\documentclass[twocolumn]{aastex631}

\usepackage{xspace}
\usepackage{upgreek}                                                                                  
\newcommand{\um}{$\upmu$m\xspace}

\newcommand{\drp}{\texttt{mirar}\xspace}
\newcommand{\wdrp}{\texttt{winterdrp}\xspace}
\newcommand{\wintertoo}{\texttt{wintertoo}\xspace}
\newcommand{\winterapi}{\texttt{winterapi}\xspace}
\newcommand{\python}{\texttt{python}\xspace}
\newcommand{\skyportal}{\texttt{SkyPortal}\xspace}
\newcommand{\pandas}{\texttt{pandas}\xspace}
\newcommand{\winter}{WINTER\xspace}

\newcommand{\kafka}{\texttt{Kafka}\xspace}
\newcommand{\avro}{\texttt{Avro}\xspace}
\newcommand{\sextractor}{\texttt{SExtractor}\xspace}

\newcommand{\datablock}{\textit{DataBlock}\xspace}
\newcommand{\datablocks}{\textit{DataBlocks}\xspace}
\newcommand{\databatch}{\textit{DataBatch}\xspace}
\newcommand{\databatches}{\textit{DataBatches}\xspace}
\newcommand{\data}{\textit{DataSet}\xspace}

\newcommand{\processor}{\textit{Processor}\xspace}
\newcommand{\processors}{\textit{Processors}\xspace}
\newcommand{\imageprocessors}{\textit{ImageProcessors}\xspace}
\newcommand{\sourcegenerator}{\textit{SourceGenerator}\xspace}
\newcommand{\sourceprocessors}{\textit{SourceProcessors}\xspace}
\newcommand{\pipeline}{\textit{Pipeline}\xspace}

\newcommand{\image}{\textit{Image}\xspace}
\newcommand{\images}{\textit{Images}\xspace}
\newcommand{\sourcetable}{\textit{SourceTable}\xspace}
\newcommand{\sourcetables}{\textit{SourceTables}\xspace}

\begin{document}

\title{A data processing pipeline for the \winter\ near-infrared surveyor using the \drp\ framework}

\author[0000-0003-2758-159X]{Viraj Karambelkar}
\email{vk2588@columbia.edu}
\altaffiliation{NASA Hubble Fellow}
\affiliation{Columbia University, 538 West 120th Street 704, MC 5255, New York, NY 10027}

\author[0000-0003-2434-0387]{Robert Stein}
\email{rdstein@umd.edu}
\altaffiliation{Neil Gehrels Prize Postdoctoral Fellow}
\affiliation{Department of Astronomy, University of Maryland, College Park, MD 20742, USA}
\affiliation{Joint Space-Science Institute, University of Maryland, College Park, MD 20742, USA} 
\affiliation{Astrophysics Science Division, NASA Goddard Space Flight Center, Mail Code 661, Greenbelt, MD 20771, USA} 
\collaboration{20}{These Authors Contributed Equally To This Work}

\author[0000-0002-7197-9004]{Danielle Frostig} 
\affil{Center for Astrophysics $\vert$ Harvard \& Smithsonian, 60 Garden Street, Cambridge, MA 02138, USA}

\author[0000-0002-3841-380X]{Saarah~Hall}
\affiliation{Department of Physics and Astronomy, Northwestern University, 2145 Sheridan Road, Evanston, IL 60208, USA}
\affiliation{Center for Interdisciplinary Exploration and Research in Astrophysics (CIERA), 1800 Sherman Ave., Evanston, IL 60201, USA}

\author[0000-0002-5619-4938]{Mansi M. Kasliwal}
\affiliation{Cahill Center for Astrophysics, California Institute of Technology, Pasadena, CA 91125, USA}

\author[0000-0002-2184-6430]{Tomás Ahumada}
\affiliation{Cerro Tololo Inter-American Observatory/NSF NOIRLab, Casilla 603, La Serena, Chile}
\affiliation{Cahill Center for Astrophysics, California Institute of Technology, Pasadena, CA 91125, USA}

\author[0000-0002-8262-2924]{Michael W. Coughlin}
\affiliation{School of Physics and Astronomy, University of Minnesota, Minneapolis, MN 55414}

\author[0009-0008-3603-0013]{Thomas Culino} 
\affiliation{Cahill Center for Astrophysics, California Institute of Technology, Pasadena, CA 91125, USA}

\author[0000-0002-8989-0542]{Kishalay De}
\affiliation{Columbia University, 538 West 120th Street 704, MC 5255, New York, NY 10027}
\affiliation{Center for Computational Astrophysics, Flatiron Research Institute, 162, 5th Ave, New York, NY 10010}

\author[0009-0001-7793-3680]{Sulekha Kishore}
\affiliation{Institute for Data, Systems, and Society,
Massachusetts Institute of Technology, Cambridge, MA 02139, USA}

\author[0009-0003-6181-4526]{Theophile Jegou du Laz}
\affiliation{Cahill Center for Astrophysics, California Institute of Technology, Pasadena, CA 91125, USA}

\author[0000-0002-4585-9981]{Nathan P. Lourie}
\affiliation{MIT-Kavli Institute for Astrophysics and Space Research, 77 Massachusetts Ave., Cambridge, MA 02139, USA}

\author[0000-0001-6331-112X]{Geoffrey Mo}
\affiliation{Cahill Center for Astrophysics, California Institute of Technology, Pasadena, CA 91125, USA}
\affiliation{Observatories of the Carnegie Institution for Science, 813 Santa Barbara Street, Pasadena, CA 91101, USA}

\author[0009-0001-4683-388X]{Tanishk Mohan}
\affiliation{Department of Physics, Indian Institute of Technology Bombay, Powai, Mumbai 400076, India}

\author[0000-0003-4725-4481]{Sam Rose}
\affiliation{Cahill Center for Astrophysics, California Institute of Technology, Pasadena, CA 91125, USA}

\author[0000-0002-9453-7735]{Benjamin R. Roulston}
\affiliation{Department of Physics, Clarkson University, 8 Clarkson Ave, Potsdam, NY 13699, USA}
\affiliation{Institute for STEM Education, Clarkson University, 8 Clarkson Ave, Potsdam, NY 13699, USA}

\author[0009-0005-2987-0688]{Aditya Pawan Saikia}
\affiliation{Department of Physics, Indian Institute of Technology Bombay, Powai, Mumbai 400076, India}

\author{Mallika Sheshadri}
\affiliation{Department of Physics and Astronomy, Northwestern University, 2145 Sheridan Road, Evanston, IL 60208, USA}

\author[0000-0003-3769-9559]{Robert A. Simcoe}
\affiliation{MIT-Kavli Institute for Astrophysics and Space Research, 77 Massachusetts Ave., Cambridge, MA 02139, USA}

\author[0000-0001-9226-4043]{Jamie Soon}
\affiliation{Research School of Astronomy and Astrophysics, Australian National University, Cotter Rd, Weston Creek ACT 2611, Australia}

\author[0009-0005-8230-030X]{Aswin Suresh}
\affiliation{Department of Physics and Astronomy, Northwestern University, 2145 Sheridan Road, Evanston, IL 60208, USA}
\affiliation{Center for Interdisciplinary Exploration and Research in Astrophysics (CIERA), 1800 Sherman Ave., Evanston, IL 60201, USA}



\begin{abstract}
We present the data reduction and transient detection pipeline for the Wide-field Infrared Transient Explorer (WINTER) surveyor and report its on-sky performance. The WINTER camera utilizes cost-effective InGaAs sensors as alternatives to traditional IR sensors, and is mounted on a dedicated 1-m robotic telescope at Palomar Observatory. The WINTER camera has six detectors producing a combined field-of-view of 1.2\,sq.\,deg. equipped with y, J, and shortened-H bands. WINTER saw first light in June 2023 and has been operating robotically since. The WINTER data processing pipeline (\wdrp) has been implemented within the broader framework \drp : a modular, open-source \texttt{python} package developed for realtime processing of images from time-domain surveys. \wdrp  performs end-to-end data processing implementing data reduction and image subtraction to go from raw dithered WINTER images to transient alerts in the \avro format, which are then sent to \skyportal for vetting and follow-up. During a year of observations in 2024, \winter achieved \emph{J-}band median 5-$\sigma$ depths ranging from 18.1-18.8\,mag (AB) on its six detectors in 960\,second integrations as part of its survey, with an astrometric accuracy of $\approx0.2$\,arcsec (a fifth of a pixel) and a detector-performance limited photometric accuracy ranging from $\approx0.09-0.18$\,mag for its six detectors. We present early science results from \winter, which include the identification of a stellar merger in M31, dust-enshrouded outbursting young stellar objects and classical novae in the Galactic plane, NIR followup of known supernovae, and multi-messenger follow-up of neutrinos, gravitational waves, fast X-ray transients and gamma-ray bursts.  

\end{abstract}



\section{Introduction} \label{sec:intro}

Time-domain astronomy is in an era of  `big data'. The last decade has seen the advent of several automated wide-field optical surveys that scan large areas of the sky every night to search for transient cosmic explosions such as the Zwicky Transient Facility (ZTF; \citealt{ztf_bellm_19}), the Asteroid Terrestrial-impact Last Alert System (ATLAS; \citealt{Tonry2018}), the All-Sky Automated Survey for Supernovae (ASAS-SN; \citealt{Shappee2014}), PanSTARRS \citep{ps1}, BlackGem \citep{Groot2024}, The Gravitational-wave Optical Transient Observer (GOTO, \citealt{Steeghs2022}), and the Large Array Survey Telescope (LAST, \citealt{Ofek2023_last}). The large data volume generated by these surveys has required the development of an infrastructure ecosystem to support real-time processing of astronomical images, image subtraction relative to reference templates to identify candidate transient alerts \citep{zogy, Hu2022_sfft, Becker2015}, and rapid disbursements of these alerts to the wider astronomy community for follow-up studies \citep[e.g.][]{skyportal_23, Nordin2019_ampel}. The new Legacy Survey of Space and Time (LSST) with the Vera C. Rubin Observatory represents another leap forward \citep{Rubin}, providing deep optical survey data that is over an order of magnitude larger than previous surveys. 

Unlike rapid advancements in optical time-domain surveys, the dynamic infrared (IR) sky has remained largely unexplored with far fewer IR time-domain surveys \citep{Kasliwal2019BAAS}. This is primarily because of the substantially larger cost of IR detectors compared to optical devices of similar size. Most historical IR surveys have targeted small areas of the sky. The Spitzer Infrared Intensive Transients Survey (SPIRITS, \citealt{Kasliwal2017ApJ}) surveyed $\sim100$ nearby galaxies at 3.6 and 4.5\,\um and uncovered a plethora of dusty IR transients that did not have any optical counterparts \citep{Jencson2019_spirits}. The VISTA Variables in the Via Lactae (VVV, \citealt{Minniti2010}) and VVV extended (VVVX, \citealt{Saito2024}) surveys targeted regions towards the Galactic bulge and mid-plane from 2009 to 2023 at near-IR (NIR) wavebands to characterize transients and variable stars in the dustiest regions of the Milky Way. The Supernovae UNmasked By Infra-Red Detection (SUNBIRD, \citealt{Kool2018}) survey used adaptive optics on the Gemini South telescope to identify supernovae (SNe) in the dense, dusty nuclei of select ultra-luminous infrared galaxies. 

The first all-sky, high-cadence exploration of the dynamic IR sky was provided by the Palomar Gattini-IR (PGIR, \citealt{gattini}) surveyor, which has been surveying the entire northern sky at a cadence of $\approx2-3$ days with its 25\,sq.\,deg. \emph{J-}band camera mounted on a 30\,cm telescope since 2018. PGIR reached a median depth of m$_{J}=$15.7\,mag (AB), and has uncovered hidden populations of dusty Galactic novae \citep{De2021} and variable stars \citep{Karambelkar2021rcb, Karambelkar2024rcb, Suresh2024}. In addition to PGIR, a recent pipeline to identify mid-IR transients in the decade-long observations from the NEOWISE reactivation of the WISE satellite surveyor \citep{Mainzer2014} has discovered a large number of mid-IR transients over the entire sky \citep{De2024_m31}. These studies have demonstrated the immense potential of IR surveys in identifying large populations of dusty cosmic explosions that are missed by optical surveys, and have highlighted the need for further systematic IR time-domain exploration. 

In addition to dusty transients, IR surveys can be particularly effective for electromagnetic followup of gravitational waves (EMGW). The kilonovae associated with binary neutron star and neutron star black hole mergers are expected to exhibit viewing-angle-dependent blue emission that fades rapidly ($<$1 week), while the NIR emission is predicted to be more isotropic and longer-lived \citep{Kasen:2013xka, Kasen:2017sxr, Metzger:2019zeh}. This was confirmed for GW\,170817, which had significantly longer-lived IR emission compared to the optical \citep[see e.g.][]{Tanvir:2017}. Thus kilonovae are more detectable in the near-infrared than the optical \citep{Zhu:2020ffa, Frostig:2022}, yet EMGW followup has primarily focused on optical wavelengths. The benefits of IR surveys have fueled efforts to build ground-based wide-field IR surveyors in the last few years, such as the PRime Focus Infrared Microlensing Experiment (PRIME, \citealt{Konndo2023, Sumi_2025}), and Wide-field Infrared Transient Explorer (\winter, \citealt{Lourie2020, Frostig2024, Frostig2025}).

\winter is a new NIR camera installed on a dedicated 1\,m robotic telescope at Palomar Observatory in Southern California. The \winter camera has a field of view of 1.2\,sq\,deg, and employs novel indium-gallium-arsenic (InGaAs) sensors as a cheaper alternative to traditional IR detectors (e.g., \citealt{Simcoe2019, Pedersen2024}). WINTER is equipped with y, J, and shortened-H (Hs) filters, covering the 1-1.7\,\um wavelength range. WINTER was designed for EMGW followup during the International Gravitational Wave Network's O4 observing run. Additionally, WINTER's wide FOV and NIR sensitivity enables it to conduct systematic searches for dusty NIR transients. \winter saw first light in June 2023 and has been operating robotically since. The new sensor technology and IR scientific landscape probed by \winter presents unique challenges for realtime data processing. 

Here we describe the data reduction and transient detection pipeline for \winter and report on the system performance of the InGaAs camera. This pipeline is implemented in an open-source modular python framework (\drp) that has been developed for real time processing of images from time-domain surveys and follow-up instruments. The paper is organized as follows: Section \ref{sec:infrastructure} describes the WINTER observatory and a dedicated API for interacting with the observatory. Section \ref{sec:drp} introduces the \drp data reduction framework, and Section \ref{sec:winterdrp} outlines the exact pipeline used to process \winter\ images. Section \ref{sec:performance} outlines \winter's on-sky performance and Section \ref{sec:early_results} presents early science results from \winter. We conclude with a summary of the lessons learned from \winter in Section \ref{sec:summary}.

\section{The WINTER observatory}
\label{sec:infrastructure}
\begin{figure*}
    \centering
    \includegraphics[width=0.5\textwidth]{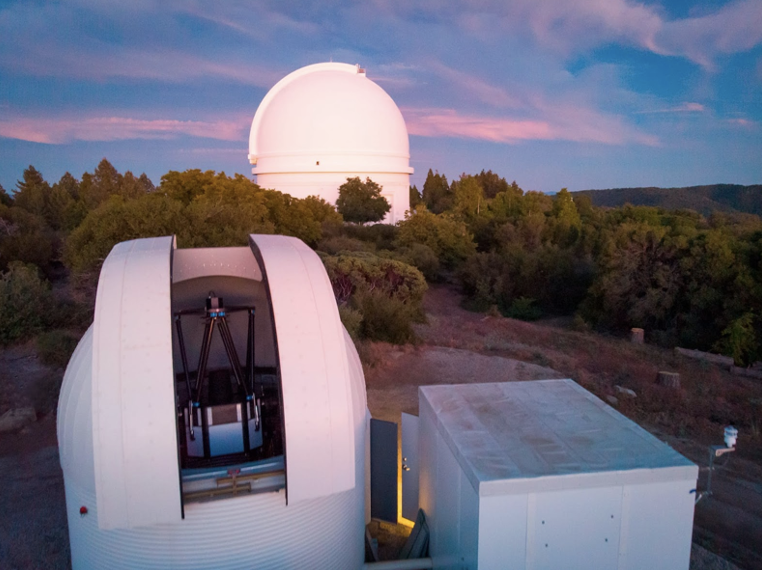}\includegraphics[width=0.5\textwidth]{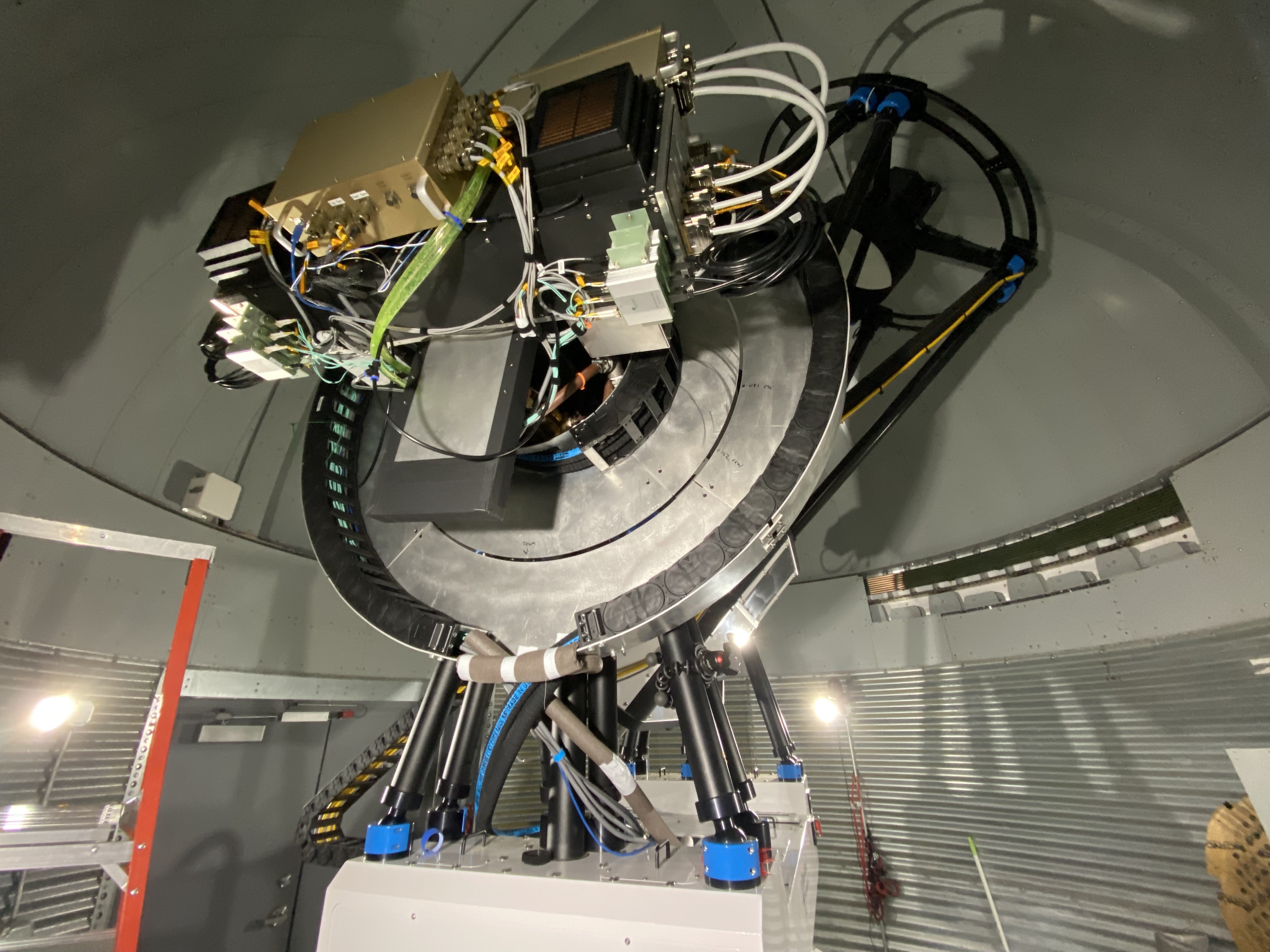}
    \caption{\emph{Left:} A picture of the WINTER dome and 1\,m. telescope, with the 200-inch Hale telescope in the background (credit: Matt Dieterich/\url{www.MattDieterich.com}). \emph{Right: } The WINTER camera mounted on the 1\,m. telescope.}
    \label{fig:winter_telescope}
\end{figure*}

The WINTER camera is mounted on a dedicated 1\,m telescope at Palomar Observatory, and consists of six infrared detectors featuring novel InGaAs sensor technology \citep{Frostig:2022SPIE, Malonis:2020}, a novel fly's-eye optical design \citep{Lourie2020}, and custom opto-mechanics \citep{Hinrichsen:2020}. Further technical details on the WINTER camera are presented in \citet{Frostig2025}. The six InGaAs detectors together produce a field-of-view of 1.2\, deg. x 1\,deg with a pixel scale of 1.13\arcsec\xspace per pixel. Figure \ref{fig:winter_telescope} shows the dome, telescope, and WINTER camera at Palomar Observatory. The telescope has been operating completely robotically since June 2023 and has been executing an observational campaign comprising surveys motivated by specific science cases and target-of-opportunity (ToO) observations. 

\section{Data Reduction Framework - \drp} \label{sec:drp}
A data reduction framework has been developed that was explicitly designed to be modular, flexible and telescope-agnostic. The Modular Image Reduction and Analysis Resource (\drp) is an open-source code built using \texttt{python}, and is available on \texttt{Github}\footnote{\url{http://www.github.com/winter-telescope/mirar}} and \texttt{PyPI}. It has extensive unit tests to ensure reproducibility, and relies on systematic type hinting to reduce end-user errors. \drp\ is built around a series of discrete configurable operational units known as \processors, which act in sequence on data.

\begin{figure*}
    \centering
    \includegraphics[width=\textwidth]{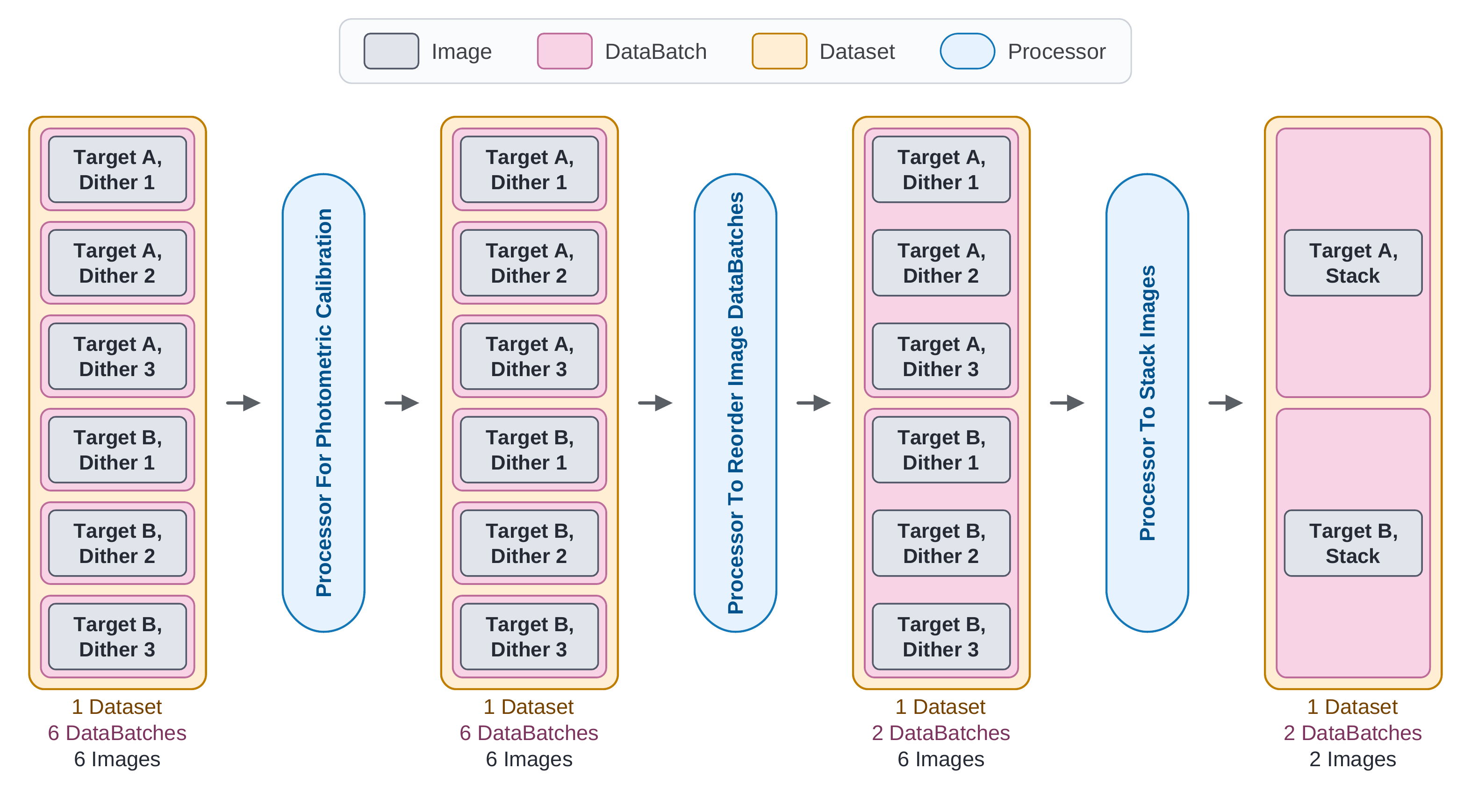}
    \caption{Illustration of the data structure usage in \drp. In the above example, a dataset initially consists of six images. These correspond to three images of Target A, and another three images of Target B. The images begin grouped into six separate batches, and undergo photometric calibration. The images are then reorganised into two batches (one batch per target). Finally, each batch is stacked together, yielding one stacked image per batch.}
    \label{fig:datastructure}
\end{figure*}

\subsection{\drp Data Structure}

\drp depends on handling data with a uniform format. Each \processor receives groups of standardized data known as \datablocks, performs a specific operation on that data, and then returns a \datablock. Several key fields are required by \drp to be in every \datablock, and these fields are added as raw images are read in. These include a full processing history, which is continuously updated as data passes through processors, and other vital metadata such as raw image path and time of exposure. There are two basic types of \datablock:

\begin{enumerate}
    \item \image: An \image object contains image data in a numpy array \citep{numpy}, as well as an astropy `header' \citep{astropy:2013,astropy:2018,astropy:2022}. There is only one header and one 2D array per \image, so multi-extension fits files are converted to multiple individual \image objects when read in by \drp. 
    \item \sourcetable: A \sourcetable object contains a table of sources in a \pandas DataFrame format \citep{pandas}, as well as a `metadata' dictionary that is analogous to the \image header. The metadata in a \sourcetable typically contains information which is common to all sources in the \sourcetable.
\end{enumerate}

At the highest level, all data forms part of a single \data (corresponding to e.g. a whole night of data). Each \processor receives a \data as input, and returns a \data as output. Within a \data, the individual \datablocks are grouped into \databatches, each containing at least one \datablock. After receiving a \data, \processors will typically loop over each \databatch in turn. For example, a processor to stack images may receive a \databatch containing eight \images corresponding to eight dithers, and then generate a \databatch containing one \image which is the combined stack of those dithers. \processors will typically treat each \databatch within the \data independently, and can act on different \databatches simultaneously through multi-processing. However, some \processors act instead on the entire \data at once, such as the \processors used to reorganize \databatches. 

When errors are encountered, \drp will drop the entire \databatch from further processing. For this reason, \images are often repeatedly reordered into different \databatch groups during the processing sequence. For example, each \image from a dither set may each be in an individual \databatch for astrometric and photometric calibration, then reordered into a single combined \databatch before being passed to a stacking \processor. Then, even if one dither \image produces an error, the remaining \images will still be used to generate a stacked \image. An example of the data structure is illustrated in Figure \ref{fig:datastructure}.

\subsection{\drp \processors}

From the two core \datablock types (\images and \sourcetables), there are three basic types of \processor:

\begin{enumerate}
	\item \emph{Image Processors} receive \images as an input, and produce \images as an output.
	\item \emph{Source Generators} receive \images as an input, and produce \sourcetables as an output.
	\item \emph{Source Processors} receive \sourcetables as an input, and produce \sourcetables as an output.
\end{enumerate}

\begin{figure*}
    \centering
    \includegraphics[width=0.8\linewidth]{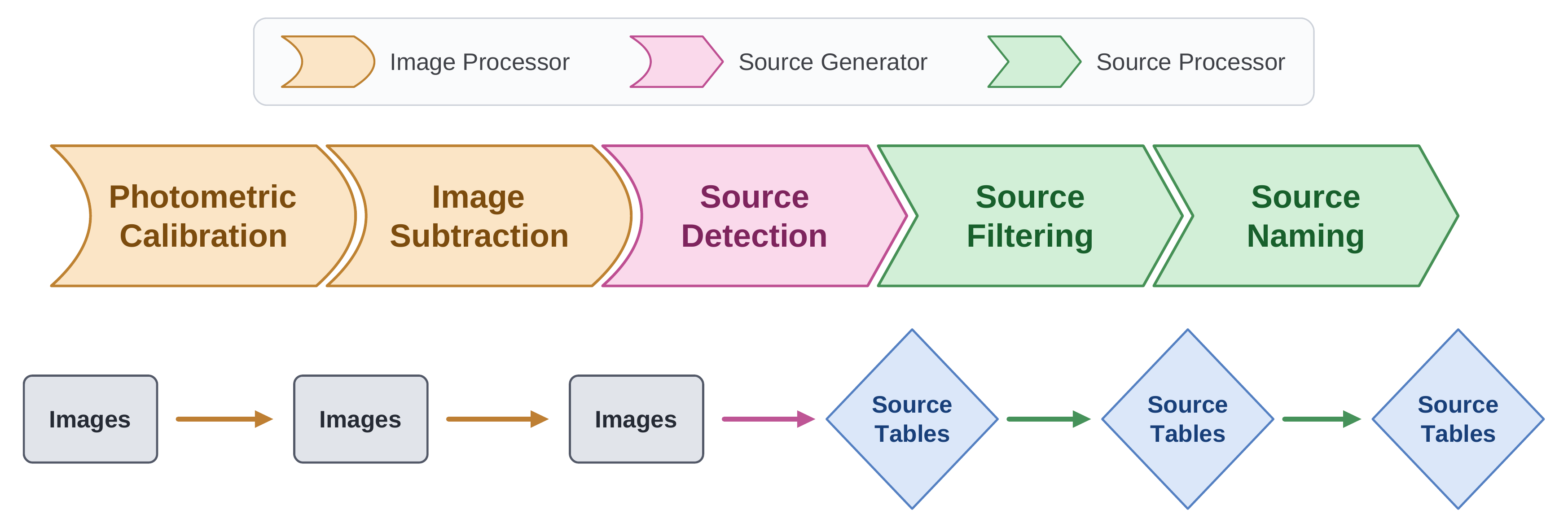}
    \caption{A illustration of a typical \processor sequence. A series of \imageprocessors are applied to \images, resulting in a set of final \images (these could be calibrated science images or subtraction images depending on the use case). A \sourcegenerator is then applied to each of these final \images, resulting in one \sourcetable for each parent \image. Finally, a series of additional \sourceprocessors are applied to the \sourcetables.}
    \label{fig:processors}
\end{figure*}

Because the data format is consistent across processors, they can be performed sequentially and in arbitrary orders. This provides substantial flexibility to recycle processors across different instruments and tasks. A typical workflow within \drp might consist of a series of image processors to detrend images, perform astrometric and photometric calibration, perform image subtraction against reference images from another survey, followed by a source detector to extract transients in the image, and finally a series of source processors to filter those detected sources and export them to external services such as a database and an alert broker. An example \processor sequence is shown in Figure \ref{fig:processors}, combining all three \processor classes.

By virtue of the modular structure in \drp, many individual \processors exist which are simply wrappers of software packages which are widely used in astronomy or industry. However, because of the uniform data standards and consistent \processor structure, it is simple to combine these different components with minimal complexity. It is similarly straightforward to add a new \processor which applies a new software package. Some prominent external libraries which are utilised within \drp \processors include:

\begin{itemize}
	\item \texttt{SExtractor} for the extraction of sources in images \citep{sextractor}.
	\item \texttt{Scamp} for providing astrometric solutions to images \citep{scamp}.
    \item \texttt{Astrometry.net} for providing astrometric solutions \citep{anet}.
	\item \texttt{SWarp} for image interpolation and stacking \citep{swarp}.
    \item \texttt{PSFex} for measuring PSFs \citep{psfex}.
	\item \texttt{ZOGY} for image subtraction against reference images \citep{zogy}.
    \item \texttt{pytorch} for machine-learning classification \citep{pytorch}.
	\item \emph{Apache} \avro for generating a serialised `alert stream' of detections.
    \item \skyportal for interacting with external `marshal' instances \citep{skyportal_19,skyportal_23}.
    \item \texttt{PhotUtils} to perform aperture/PSF photometry \citep{photutils}.
    \item \texttt{PostgreSQL} for interacting with postgres databases.
\end{itemize}

Key native \drp \processors include:

\begin{itemize}
    \item \textit{RawImageLoader} and \textit{MEFloader} for loading raw fits files and converting them to \images.
    \item \textit{CalHunter} for finding and loading missing calibration images.
    \item \textit{ImageSaver} for writing an \image to a fits file.
    \item \textit{ImageBatcher}, \textit{ImageDebatcher} and \textit{ImageRebatcher} for reorganising \images into different \databatches.
    \item \textit{ImageSelector} and \textit{ImageRejector} for removing \images that match user-specified criteria.
    \item \textit{CSVLog} for generating a csv log of \images in a \data, with user-selected columns.
    \item \textit{PhotCalibrator} for performing photometric calibration against a user-selected reference catalogue.
    \item \textit{DarkCalibrator}, \textit{BiasCalibrator} and \textit{FlatCalibrator} for respectively performing dark-subtraction, bias-subtraction and flat correction. 
    \item \textit{CustomImageBatchModifier} as a catch-all for applying any user-provided python function to \databatches.
    \item \textit{SourceExporter}, \textit{JsonExporter}, \textit{CSVExporter} and \textit{ParquetExporter} for exporting \sourcetables to pickle, JSON, CSV or Parquet files respectively.
    \item \textit{SourceLoader}, \textit{JsonLoader}, \textit{CSVLoader} and \textit{ParquetLoader} for loading \sourcetables from pickle, JSON, CSV or Parquet files respectively.
\end{itemize}

\subsection{\drp pipelines and configurations}

While users can analyze data with any arbitrary series of processors, \drp contains a series of tailored sequences for selected instruments. A \pipeline class is created for each new instrument, containing all of the telescope-specific information that is needed for data processing. Most importantly, each \pipeline  has a function to read in raw data from the corresponding telescope, and ensure that the image and header are formatted correctly for the \imageprocessors. In this way, multiple different telescopes can use the same sequence of processors to reduce images. 

A single chain of \processors is known as a `configuration', and a given \pipeline may have several pre-built configurations to choose from. For example, one configuration might simply generate a CSV log for a particular night of data, while another configuration completely reduces this raw data to produce difference images, and a third might reduce the data without subtraction and produce lists of detected sources. Each configuration must have a name, and can be selected based on the unique combination of instrument and configuration. In addition to the aforementioned pipeline for \winter, \drp also has already-implemented pipelines for the following instruments:

\begin{itemize}
    \item Wide Field Infrared Camera (WIRC) on the Palomar 200-inch telescope \citep{wirc}.
    \item Wafer-Scale Imager for Prime (WaSP) on the Palomar 200-inch telescope.
    \item Gemini Multi-Object Spectrographs (GMOS) on the Gemini-North telescope \citep{gmosn}.
    \item Super PalomaR INGaas (SPRING) Camera on the 1-m Palomar Telescope \citep{Frostig2025}
    \item Large Monolithic Imager (LMI) on the 4-m Lowell Discovery Telescope \citep{ldt_12,ldt_14}
    \item Spectral Energy Distribution Machine V2 (SEDMv2) optical camera on the 2.1m Kitt Peak telescope. 
\end{itemize}

There are also plans for the pipeline of the Dynamic REd All-sky Monitoring Survey (DREAMS) IR telescope \citep{Soon2022_dreams} to use \drp.

\subsection{\drp execution and monitoring}

Pipelines can be run via the command line on a particular night of data, in which case that night's data will be processed sequentially. Calibration image requirements can be specified in advance, and if any of these are missing within a night's data, the \emph{CalHunter} processor can be used to search for suitable calibration images in previous nights. \drp\ uses python's inbuilt \texttt{logger} class throughout, so that output at a chosen verbosity can be printed to the terminal or saved to a file. In addition, \drp\ has systematic error catching, handling and aggregation, so any errors raised during processing will be tracked. After processing has been completed, a dedicated error log can be produced which groups errors raised by type, and aggregates the list of images for which that error was raised. This feature is useful for gathering statistics to understand where processing might be failing.

In addition to the `offline' mode, \drp\ has a \emph{Monitor} class for real-time processing. This mode can also be called from the command line, and uses the python \texttt{watchdog} module to watch a specified directory for new images. After checking to ensure the images are complete (i.e they are not partially transferred), the monitor will then run the image through a specified real-time pipeline. After a chosen number of hours, a post-processing step can be run to provide an email summary of the processing. Typically, a new monitor instance would be created each afternoon to process the upcoming night of observation data, with the post-processing summary step then run the next morning when no new images are expected. The monitor class can persist for a user-selected duration, which could be several days, to ensure that even if data is delayed it can still ultimately be processed.   

\subsection{Documentation}
\drp is documented through in-code comments and docstrings, which are rendered by ReadTheDocs\footnote{\url{https://mirar.readthedocs.io/en/latest/}}. These are augmented by dedicated documentation pages describing high-level information and installation instructions. Pipeline configurations are automatically rendered as a flowchart for visualization, and each flowchart is included in automatically-generated documentation pages for that instrument. Users can browse available instruments\footnote{\url{https://mirar.readthedocs.io/en/latest/autogen/pipelines.html}}, and all available configurations for these instruments. 

\section{The \winter\ Pipeline (\wdrp)}
\label{sec:winterdrp}

\begin{figure*}
    \centering
    \includegraphics[width=\textwidth,height=\textheight,keepaspectratio]{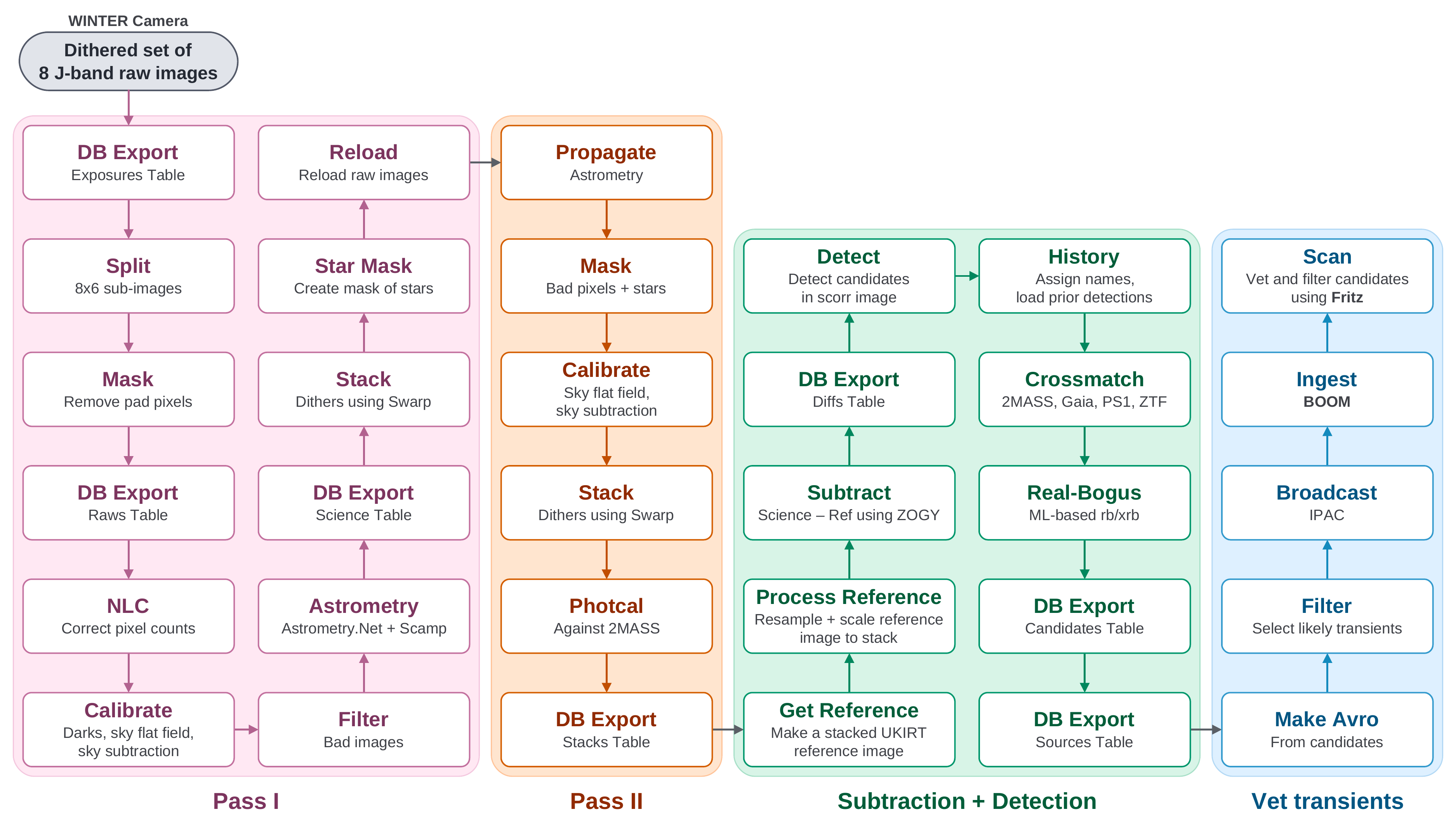}
	\caption{Flow chart of \wdrp.}
	\label{fig:winterflow}
\end{figure*}

In this section, we describe the \wdrp flow of data processing for \winter, from raw photoelectrons to uploaded transient candidates in \skyportal. The flow is also shown in Figure \ref{fig:winterflow}.

\subsection{Raw \winter data}
\begin{figure*}[hbt]
    \centering
    \includegraphics[width=\textwidth]{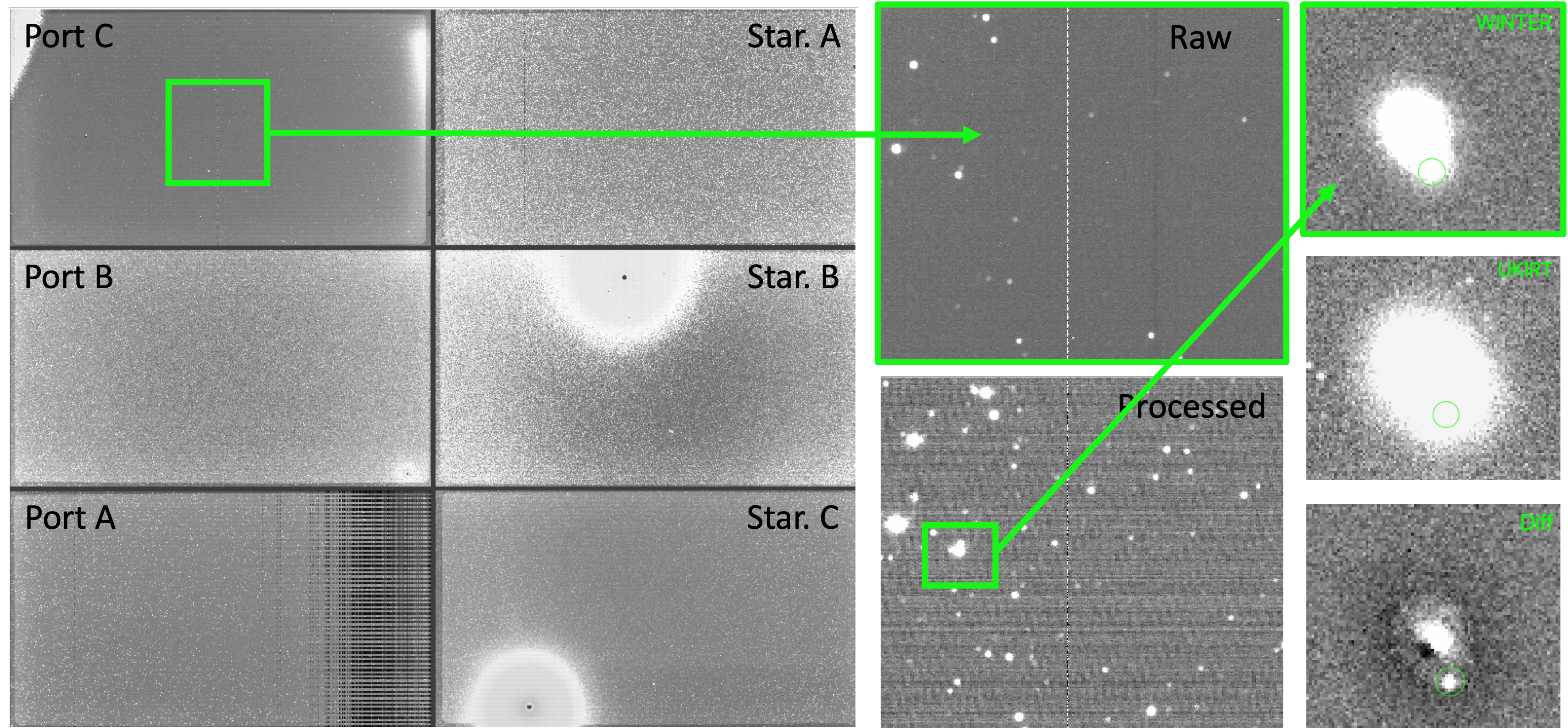}
    \caption{An example of a raw WINTER image as it is processed through different stages of the pipeline. The multi-extension fits (MEF) file with all six labelled sensors is shown in the \emph{left} panel, and the smaller images show zoom-ins of the raw, processed, reference, and difference images around a transient. The big white spots on the Star B and Star C sensors mark glow spots. The Star B sensor was replaced in April 2024 with a new sensor that does not have the glow spot, but has one nonfunctional output of its eight readout channels.}
    \label{fig:raw_to_diff}
\end{figure*}
Observations are coordinated by the \winter scheduler \citep{Frostig2025}, with user discretion in terms of exposure time and dither sequencing. However, for both routine survey observations and ToO observations with \winter, observations are conducted with exposure times of 120s in the \emph{J}-band. Our standard \emph{J-}band sequence uses 8 dithers (16 minutes), used uniformly for all surveys, while our follow-up observations occasionally use up to 30 dithers (60 minutes) for deep imaging. The single dither exposure time was chosen to keep the counts in each exposure close to the well center ($\approx20000$ counts). The number of dithers is chosen based on the total integration time required to achieve the targeted depth. The dither distance is also user configurable, but based on the physical scale of detector imperfections almost all observations are conducted with our recommended 90\arcsec dither step. 

Images are acquired using using custom sensor software and firmware on the export-controlled \emph{Freya} computer, physically located in the \winter control room at Palomar Observatory. The resultant exposures are sent to the neighboring non-export controlled \emph{Odin} machine, which is more widely accessible. Raw WINTER exposures are initially saved as six-frame  multi-extension fits files (one frame per WINTER sensor). These images are then transferred from \emph{Odin} to Caltech using the High Performance Wireless Research and Education Network (HPWREN) via an \texttt{rsync} command that runs every minute. The images are copied to the dedicated \emph{\winter} server at Caltech, with raw data organized into a separate directory for each night.

After raw images arrive, all subsequent processing of data is done with the realtime \winter\ pipeline (\wdrp), built using the \drp\ framework. A \wdrp\ \texttt{monitor} is started each day at 1AM Pacific, and is assigned to watch the directory for that night's data. As images appear on the dedicated \winter\ machine at Caltech, they are assigned to one of 40 available multi-processing threads. The images are grouped by dither set, and then processed as a single \databatch. 

Every raw image is a multi-extension fits file with each extension corresponding to one of the six detectors, identified as Port A/B/C or Starboard (`Star') A/B/C (see Figure \ref{fig:raw_to_diff}). The Star C and Star B detectors have unusable areas due to glow spots. The Star B detector was replaced in April 2024 with a new engineering grade sensor that does not have glow spots, but one of its eight readout channels is nonfunctional leading to an eighth of its pixels, dispersed over the full detector, being non-functional \citep{Frostig2025}. Glow spots are also visible on Port C, but they are limited to the detector edges, while Port A has a large number of dead pixels towards its right edge. The detectors also have several ``hot" pixels that appear to be transient, with large differences in pixel values in subsequent images. We attempt to mitigate the effect of these detector defects through aggressive masking (see Section\,\ref{sec:masking}) and dithering. Given the different outage patterns, sensitivities, and dark current levels of each detector, the \wdrp pipeline processes each detector independently. Figure \ref{fig:raw_to_diff} shows an example of a raw image as it goes through all steps of the pipeline to identify a transient in the image. The next subsections describe specifics of the steps involved. 

\subsection{Non-linearity corrections}
\label{sec:nlc}
WINTER’s detectors exhibit a higher degree of nonlinearity than initially expected. The strongest deviations appear at lower signal levels (9000 -- 14000 counts), but significant deviations persist in the most linear region (28000 to 44000 counts), well before the expected high-end nonlinearity near saturation. To correct for these effects, a piecewise polynomial calibration function is applied to each pixel, approximating a linear response. More details about \winter non-linearity and derivation of the polynomial coefficients are described in \citet{Frostig2025}, and follows the method from \citet{layden_25}. Briefly, for a particular hardware configuration, a series of increasingly bright flats are taken to measure the pixel response, and these data are used to fit the underlying curve parameters. These data were taken with the fully integrated instrument by taking dome flats at increasingly long exposure times, using an incandescent light bulb as a source. Having measured these parameters, any raw count value can then be converted back to a `linearized' count value. Here, we describe the pipeline implementation of these nonlinearity corrections. 

A dedicated Python package, \texttt{winternlc}\footnote{\url{https://github.com/winter-telescope/winternlc}}, was created to implement these non-linearity corrections. The parameter maps are stored as six-frame fits files, and these are downloaded locally when the package is run. Crucially, the parameters themselves depend on the exact settings of the hardware and firmware, meaning that they change over time as settings are adjusted. The package therefore has strict versioning and includes multiple versions of the correction maps corresponding to different hardware states. The non-linearity correction is applied in \drp to every individual \winter image, including both calibration and science images.

\subsection{Dark subtraction}
\label{sec:dark_calibration}
\begin{figure}[hbt]
    \centering
    \includegraphics[width=0.5\textwidth]{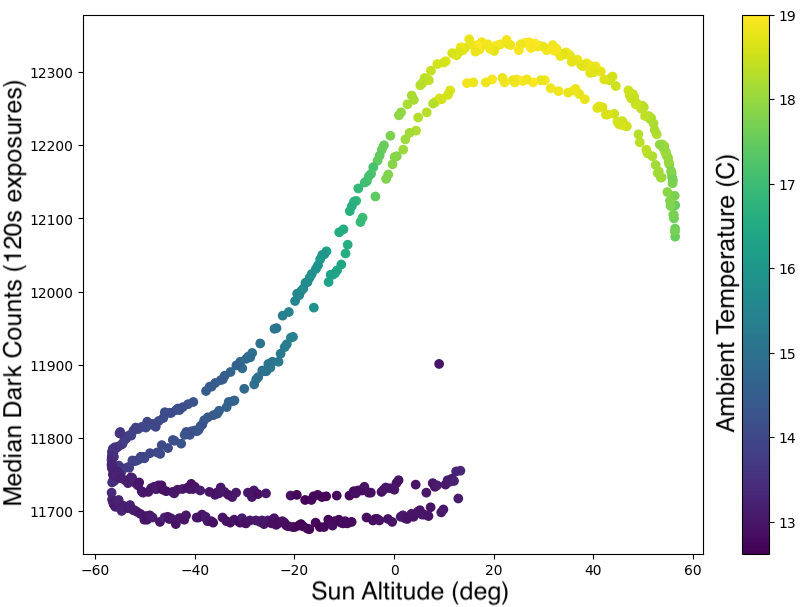}
    \caption{Median counts in 120\,s dark frames from the Port C detector taken in a continuous eighteen hour timespan from noon to sunrise the next day, plotted as a function of sun altitude. The points are color-coded by the ambient temperature. The dark counts vary substantially with sun altitude and ambient temperature. WINTER uses dark frames taken at midnight every day to reduce the effect of this. Additionally, the dark counts randomly jump between two distinct curves. This variation remains undiagnosed, and sets a noise floor for the WINTER images.}
    \label{fig:dark_counts_variation}
\end{figure}

The next step in the calibration routine is dark subtraction. The operating temperatures of the on-sky WINTER detectors range from -30 $\mathrm{^{o}}$C to -15 $\mathrm{^{o}}$C, substantially warmer than the design requirement of -50 $\mathrm{^{o}}$C (see \citealt{Frostig2024} for possible causes). As a result, the dark current of WINTER is comparable to the sky background counts. Furthermore, the measured dark counts appear to be unstable and correlate with several environmental parameters. Figure \ref{fig:dark_counts_variation} shows the variation of counts in 120 sec dark exposures taken over an 18 hour period starting from noon local time to sunrise of next day. Dark images taken during the day have substantially larger counts (often exceeding the dark+sky counts of images taken during the night) pointing to a light leak in the optical system. To mitigate the effect of the light leak, \wdrp uses darks taken at midnight. The dark counts also correlate strongly with the ambient temperature, suggesting that the detectors are sensitive to thermal emission from components in the optical system. Similar variations have been reported on the SPECULOOS camera that also utilizes InGaAs sensors for NIR astronomical imaging and photometry of exoplanet transits \citep{Pedersen2024}. To mitigate night-to-night temperature variations, dark images are acquired on a nightly basis. This strategy does not correct for temperature variations within a night, though we find that these variations are substantial from sunset to midnight, but relatively stable from midnight to sunrise. Additionally, we observe a third effect, where even at constant temperature the dark current fluctuates between two possible levels, separated by $\approx50-100$ counts (see Figure \ref{fig:dark_counts_variation}). This effect is currently undiagnosed and we do not have a way of correcting for it. Together, these variations in the dark current of $\sim5\%$ relative to sky counts contribute to the systematic uncertainty floor for WINTER photometric measurements (Section\,\ref{sec:performance_phot}). 

The sensitivity of the dark counts to the ambient temperature traces back to the \winter camera design. \winter dark frames are acquired with the detector viewing a first-surface mirror through reimaging optics. The mirror occupies one slot in the filter tray at the entrance to the WINTER camera and is inserted into the beam automatically during dark image acquisition. The mirror's low emissivity limits the thermal flux it emits, reducing the optical loading on the detectors, but the residual mirror emission still contributes to the measured dark signal, as do the reimaging optics and the detector housing. These contributions are difficult to disambiguate, though in principle they could be separated given sufficient measurements of component temperatures. The 1.7\,\um\ cutoff of InGaAs makes it less susceptible to thermal loading than the 2.5\,\um\ cutoff HgCdTe typically used in NIR instruments, but the dark loading would be better controlled in future iterations by a design incorporating active cooling or thermal control of the instrument enclosure, improved thermal shielding within the detector package, and a cooled aperture stop.

In summary, dark frames are taken every night at midnight local time. Five sets of dark exposures for all integration times of science exposures for the night (typically 120 sec. for \emph{J-}band observations) are acquired and median stacked to generate a master dark that is subtracted from the raw exposures. On a clear night, between 2000-3000 sky counts are detected in the dark subtracted images. Dark subtracted images with counts lower than 500 are marked as bad and are dropped from the pipeline.

\subsection{Masking}
\label{sec:masking}
We aggressively mask areas around glow spots on all detectors. Additionally, we construct a bad pixel mask using dithered frames for each exposure by identifying pixels that are 3$\sigma$ outliers in every dithered exposure from that observation. We find that this strategy identifies the majority of the hot and dead pixels in the image. These bad pixels constitute $\approx10\%$ of the entire detector. In addition to these, we also find a small number of transient bad pixels that are present in some but not all images. These are possibly correlated with the undiagnosed fluctuations in the dark current. These pixels seem to appear randomly in some dithers, and amount to $\sim1\%$ of the entire detector.

\subsection{Flat Calibration and sky subtraction}
The WINTER detectors have a pattern of eight interweaved readout channels, each with different (unquantified) gain values, together with non-negligible pixel-to-pixel gain variations. There is also a strong vignetting pattern across the focal plane. We attempt to calibrate for these effects through flat-fielding. We experimented with four strategies for constructing a flat field - 1) Dome flats taken using a lamp installed in the WINTER dome, 2) Sky flats taken during twilight, 3) Nightly flats constructed by stacking all science exposures from a given night, 4) Per-target nightly flats constructed by stacking all dithered science exposures of a given target. The dome flats were determined to be ineffective, as the \winter dome flats did not match the illumination and vignetting pattern seen on sky. The twilight flats had substantially larger sky counts than the majority of science images taken during the night. Non-linearity effects in the WINTER detectors (see Sec. \ref{sec:nlc}) precluded the use of twilight flats, as large residuals were visible in the science images that were flat calibrated using them, leaving nightly flats as the only viable option. This approach has been used by other infrared surveys (e.g., \citealt{gattini}), as the bright IR sky background provides a reasonable approximation of uniform illumination across the detector. We first constructed master flats by stacking all science images taken during a night and used them to calibrate the science images. However, we find a residual circular pattern across the focal plane in all calibrated images. The morphology of this pattern changes with telescope pointing and is likely due to position-dependent scattered light from the telescope. To mitigate this, instead of stacking all nightly science images, we constructed flat frames for each dither-set by stacking all dithered exposures in that observation set. This successfully removes the circular arcs and other trends from the images. However, we observe dark residual wings around bright stars and galaxies in the flat-calibrated images. To correct for this, we perform astrometry and stack these flat-calibrated images (see Sec. \ref{sec:astrometry} and \ref{sec:stacking}) and perform image segmentation on the stacked image using \sextractor to generate source masks covering the brightest sources in the image \citep{gattini}. We then go back and apply the source masks to the dark-calibrated images and stack them to regenerate sky flats, this time with bright sources masked. This strategy reduces the dark residuals around bright sources. 

Finally, a 2D background is estimated and subtracted from each flat-calibrated image  using \sextractor with a background mesh size of $128 \times 128$ pixels. 

\subsection{Astrometric Calibration}
\label{sec:astrometry}
Astrometric calibration is performed in multiple phases within \wdrp to ensure an accurate solution is found. An initial solution is obtained using \texttt{astrometry.net} \citep{anet}, using the standard Tycho 2/Gaia DR2 \citep{tycho2,gaia_crossmatch} reference index catalog stored locally on the Caltech winter machine. For this, \texttt{astrometry.net} is configured to use \sextractor to identify sources, and not to measure any distortion corrections. As a second step, \sextractor is run on the image again to generate a source catalog, which is used to calculate full astrometric solutions including distortion coefficients using \texttt{Scamp} \citep{scamp}. The joint Gaia/2MASS catalog \citep{2mass,gaia_crossmatch} is used as the reference catalog for \texttt{Scamp}. We require the reference sources to be detected with signal-to-noise ratios $>3$, not have any proper motion, and not be confused in the Gaia–2MASS cross-match (i.e., \texttt{number\_of\_mates = 0} and \texttt{number\_of\_neighbours = 1}). The astrometric accuracy (\texttt{ast\_unc}) of the solution is computed for each image by calculating the median offset between the image coordinates and the reference coordinates, and images with poor astrometric solutions at this stage ($\texttt{ast\_unc} > 0.5$\,\arcsec) are dropped from further processing. The median astrometric precision achieved for all images in year 1 of the survey (2024) was $0.21$\,\arcsec for sources detected with S/N$>10$, corresponding to about a fifth of one \winter pixel.

\subsection{Image Stacking and Photometric Calibration}
\label{sec:stacking}
The detrended and astrometry-calibrated images in a dither sequence together with their image masks are fed to the \texttt{SWarp} \citep{swarp} processor which resamples and median combines them to generate stacked images. Source catalogs for the co-added stacks are then generated using \texttt{SExtractor}, which are fed into \texttt{PSFex} \citep{psfex} to generate a point-spread-function (PSF) model that is saved to disk for downstream processing. The PSF model is then used in conjunction with a second run of \texttt{SExtractor} to generate a source catalog that includes PSF photometry for every source. This source catalog is used for photometric calibration of the coadded stack.

The augmented 2MASS point source catalog \citep{gaia_crossmatch,2mass} is used as the reference catalog for the J-band and shortened-H band images, while the PS1 \citep{ps1} catalog is used as reference for the y-band images. In all cases, the zeropoint is calculated from sources cross-matched between the stacked and reference catalogs using --

\begin{equation}
    \label{eq:zp_eqn}
    \rm{m}_{\textup{ref}} - \rm{m}_{\textup{img}} = \textup{ZP} + C \times (\textup{color}_{\text{ref}})
\end{equation}
where m$_{\rm{ref}}$ is the apparent magnitude of the source in the reference catalog, m$_{\rm{img}}$ is the instrumental magnitude measured from the \winter image, ZP is the zeropoint, C is a color coefficient, and color$_{\rm{ref}}$ is the color of the source in the reference catalog. The 2MASS $J-H$ color is used as color$_{\rm{ref}}$ for J-band images, 2MASS $H-K$ color is used for Hs band images, and PS1 $y-z$ color is used for y-band images. For every stacked image, a set of good-quality sources crossmatched between the image and reference catalogs are selected using a crossmatch radius of 1\,\arcsec and filtered further by requiring signal-to-noise threshold $>3$, full-width-half-maximum (FWHM) $< 5$\,\arcsec, apparent magnitude m$>10$\,mag, and \texttt{SExtractor} \texttt{FLAGS}=0. Images with fewer than five crossmatched sources are removed from further processing. For the remaining images, the best-fit values of ZP and C are determined from Equation\,\ref{eq:zp_eqn} using the cross-matched stars and the \texttt{optimize} module in \texttt{scipy} \citep{scipy}. 

\subsection{Difference Imaging and Transient Detection}
Difference imaging on the J and Hs-band is performed relative to public images from the United Kingdom Infrared Telescope (UKIRT) wherever available.  Most J-band reference images are from the UKIRT Hemisphere Survey (UHS, \citealt{ukirt_hemisphere}), while a small number of reference images are from the UKIRT Infrared Deep Sky Survey (UKIDSS, \citealt{Lawrence2007}) Data Release 11. The UHS images are stored locally on the \winter server at Caltech, while the UKIDSS images are downloaded in real-time from the WFCAM Science Archive \footnote{\href{http://wsa.roe.ac.uk/}{http://wsa.roe.ac.uk/}}. For y-band observations, we use reference images from PANSTARRs-1 \citep{ps1}. For every \winter image, a reference image is constructed by median-coadding all UKIRT/PS1 images that overlap with the stack \winter using \texttt{SWarp}. Before coadding, a multiplicative scaling factor is applied to the individual images to scale them to the median ZP of all images.

Coadded reference images are resampled to match the pixel scales of the science image using \texttt{SWarp} \citep{swarp}. As the native UKIRT and PS1 reference images have a finer pixel scale than \winter, resampling them to the coarser \winter pixel scale does not introduce any additional correlated noise (see e.g., \citealt{gattini}). \texttt{SExtractor} and \texttt{PSFex} are run on the resampled reference images to generate reference PSF models and reference source catalogs. The reference image is further scaled to match the sensitivity of the science image -- i.e. to ensure that common stars have the same counts in both images. Both matched images, along with their uncertainty maps, PSF models, and relative astrometric uncertainties are fed into a python implementation of the ZOGY algorithm \citep{zogy}, which yields a final `difference image', a difference uncertainty map, difference PSF model, and a significance `SCorr' image. 

Source detection is not performed directly on the difference, but rather on the corresponding SCorr image \citep{zogy}. We run \sextractor on the SCorr image to identify sources, with a peak SCorr value greater than 5. Aperture and PSF photometry is then performed at these positions in the difference image and all sources with these values are compiled in a \sourcetable. Each row of the \sourcetable also includes three `cutout images', which are 64$\times$64 pixel arrays centered on the source with image data from the parent science, reference and difference image. Metadata about the detection, for example the FWHM and the number of bad pixels in the difference cutout, are also included in the \sourcetable. 

\subsection{Crossmatching with external catalogs}
Detected sources are cross-matched to a variety of external catalogues to provide contextual information, in particular through queries first to the \texttt{kowalski} database at Caltech \citep{ztf_drb}, and lately through BOOM \citep{boom_2026}. These include crossmatches within 30" to the 2MASS point source catalogue \citep{2mass}, PS1 point source catalogue \citep{ps1}, the PS1-based star/galaxy classification catalogue \citep{tachibana_18}, and PS1 photometric redshifts \citep{ps1strm}, to the Gaia EDR3 source catalogue within 90" \citep{gaia_edr3}, as well as to ZTF-detected sources within 2\arcsec \citep{ztf_bellm_19}. Sources are also cross-matched to previous WINTER detections within 2\arcsec, providing a complete detection history of each source. At this stage, if sources have no previous detection, they are assigned a unique sequential name of the form \emph{WNTR\,25abcde}. Sources that do have previous detections will instead inherit their previous name. 

\subsection{Machine Learning Real-Bogus Classification}
\wdrp includes a dedicated `real/bogus' machine-learning stamp classifier to remove spurious detections, known as \texttt{winterrb}\footnote{\url{https://github.com/winter-telescope/winterrb}}. The design is based on the successful `deep real-bogus' (drb) classifier \citep{ztf_drb} used by ZTF, with the classifier acting directly on the three cutout image stamps. No additional metadata is required by the classifier. The model itself is built using \texttt{pytorch} \citep{pytorch}, rather than the original \texttt{TensorFlow} framework employed by ZTF, but in other respects our approach is similar. We employ a neural network with an architecture consisting of 16 layers for convolutional, pooling, batching and dropout, yielding a single final score.

The training set from the data is heterogeneous, including sources flagged by human as `real'. The vast majority of `real' sources are those crossmatched to known transients from ZTF \citep{ztf_bellm_19}. The `bogus' class is more challenging to define. While certain types of artefacts are easy to identify visually (such as a `dipole'/`yin-yang' residual, a high proper motion star or a streak), the majority of statistically significant winter candidates are observed to be single-pixel detections, and therefore suspected to be instrumental artifacts. For the purposes of training, we make the simplifying assumption that all sources not tagged as real are instead bogus. This yields a highly imbalanced dataset, consisting of 252 real detections and 184407 bogus detections.

We downselect a random sample of 20160 `bogus' detections, which we combine with the 252 `real' detections, yielding a dataset of 20412 detections with a ratio of 80 bogus to each real. We employ stratified K-fold cross validation, in which both real and bogus detections are evenly divided into 5 groups (each with $\sim$50 real and $\sim$4000 bogus).  We exploit the intrinsic symmetry of cutouts to augment our `real' sample. Each cutout can be rotated by 90, 180 or 270 degrees, and can additionally be mirrored in the x and/or y plane. This means a single detection can produce 4x2x2=16 unique permutations. Consequently, we can generate a total of 800 `real' detections for each fold from an original 50, and our training dataset thus has a final ratio of 5:1 bogus detections to real detections.

As part of K-fold cross validation, we train the classifier five times, with each permutation being tested on one of the 5 folds and trained on the remaining four groups. In this way, every one of the detections receives a score from a classifier which was not trained on it. We employ the the Adam optimizer \citep{adam} and ten training loops for each permutation to calculate average performance. Finally, for the `production' version of the classifier, we retrain once using the full dataset.

\begin{figure}
    \centering
    \includegraphics[trim=1cm 0 0 0, clip, width=0.98\linewidth]{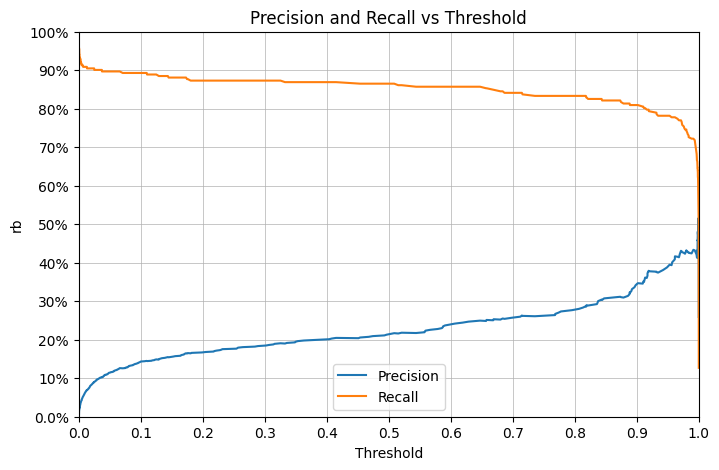}
    \includegraphics[trim=1cm 0 0 0, clip, width=0.98\linewidth]{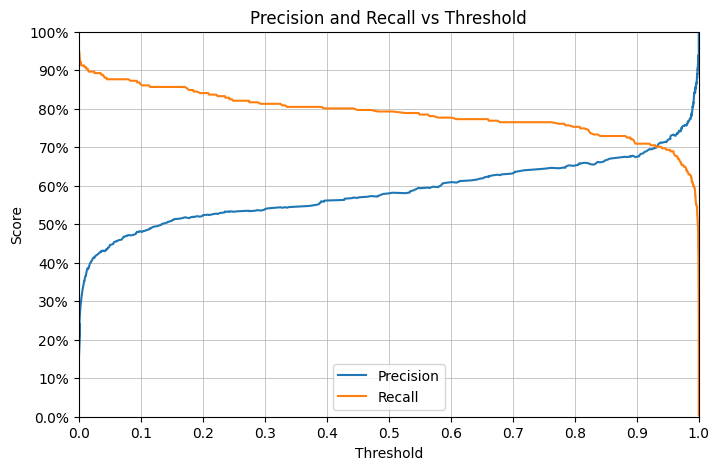}
    \caption{\textbf{Top:} Final performance of the real/bogus stamp classifier (\texttt{rb}), on the parent (unaugmented) dataset of 184659 detections. Though the precision achieved is modest ($\sim$10\% for $\sim$90\% recall), the parent dataset is extremely imbalanced (one in a thousand sources are tagged as `real'). Applying the classifier thus yields a hundred-fold increase in precision, dramatically reducing the number of bogus detections. \textbf{Bottom:} Performance of the XGBoost classifier (\texttt{xrb}), on the parent (unaugmented) dataset of 184659 detections. Combining \texttt{rb} with additional features considerably boosts performance. To maintain the same $\sim$90\% recall as the stamp classifier, precision increases substantially to $\sim$40\%, representing a further four-fold improvement relative to stamp classification alone.}
    \label{fig:winterrb}
\end{figure}

The performance of the stamp classifier is shown in Figure \ref{fig:winterrb}, with precision and recall both shown as a function of threshold score. For WINTER, our primary aim with employing a classifier to select a manageable number of number of sources in a given night for which human inspection would be feasible. Approximately one in a thousand sources in our original dataset were tagged as `real', equivalent to a precision of 0.13\%. As can be seen in Figure \ref{fig:winterrb}, applying the classifier can increase this precision to $\sim$10\% while retaining a recall of $\sim$90\%. Restated in simple terms, the stamp classifier can reject 99\% of bogus detections while retaining 90\% of real detections. 

In addition to the image cutout data, there is additional metadata available about detections which could improve real/bogus discrimination. To make use of this additional discriminating power, we train a second real/bogus classifier based on the strategy employed in \citet{tdescore}. We use XGBoost \citep{xgboost} to train a simple and interpretable decision-tree based classifier. \texttt{rb} is one variable used in this dataset, but we include several others listed in Table \ref{tab:feature_importances}. We again perform Stratified K-Fold Validation, and maintain separation between training/testing by only relying on scores from the `test' phase of the stamp classifier. As with \texttt{rb}, we generate a final production version using the complete dataset.

The performance of the XGBoost classifier (\texttt{xrb}) is shown in the lower panel of Figure \ref{fig:winterrb}. As can be seen immediately, performance is substantially improved relative to the stamp classifier alone. For the same threshold of $\sim$90\% recall, \texttt{xrb} achieves an enhanced precision of $\sim$40\%. The rejection rate is thus increased to 99.8\% of all bogus detections, and the fraction of real sources is increased three-hundred-fold relative to the parent dataset.  

For a tree-based classifier, we can directly quantify the relative importance of features. These are also listed in Table \ref{tab:feature_importances}. Beyond \texttt{rb}, which is by far the most important, additional key parameters include the number of nearby PS1 and 2MASS sources (nmtchps / nmtchtm), the distance to the nearest edge of the detector (mindtoedge) and the distance to the closest PS1-detected source (distpsnr1).

Both of the ML classifiers are deployed in production within \wdrp. The trained models were saved and exported, and can be loaded by \drp. \wdrp sequentially assigns both an \texttt{rb} value and then an \texttt{xrb} value to each new detection, for use in downstream filtering. The classifiers will continue to be periodically retrained as the training set grows, so the corresponding classifier version is given alongside each score.

\begin{table*}[htbp]
\centering
\begin{tabular}{lcl}
\hline
Feature & Importance (\%) & Description \\
\hline
\texttt{rb} & 86.0 & WINTER stamp classifier (real-bogus) score \\
nmtchps & 3.6 & Number of PS1 catalog sources matched within 30\arcsec \\
mindtoedge & 2.2 & Distance to nearest edge of the sensor \\
distpsnr1 & 1.3 & Distance to nearest PS1 source (PSF-catalog match) \\
nmtchtm & 1.1 & Number of 2MASS catalog sources matched within 30\arcsec \\
bimagerat & 1.0 & Ratio of source semi-minor axis to PSF-model semi-minor axis \\
chipsf & 0.9 & Reduced $\chi^2$ of the PSF-fit photometry \\
distgaiabright & 0.8 & Distance to nearest bright ($G<14$) Gaia source \\
SCorr & 0.7 & Peak S/N in the ZOGY score image \\
elong & 0.6 & Source elongation (a/b axis ratio) \\
aimagerat & 0.5 & Ratio of source semi-major axis to PSF-model semi-major axis \\
sumrat & 0.4 & Ratio of pixel-sum flux to PSF-fit flux \\
bimage & 0.3 & Semi-minor axis of source (\sextractor) \\
aimage & 0.3 & Semi-major axis of source (\sextractor) \\
nneg & 0.2 & Number of negative pixels in a 5$\times$5 cutout around the source \\
fwhm & 0.2 & Full width at half maximum of the detection\\
nbad & 0.0 & Number of bad pixels in a 5$\times$5 cutout around the source \\
\hline
\end{tabular}
\caption{Feature importances for the XGBoost classifier (\texttt{xrb}). Though \texttt{rb} from the stamp classifier remains by far the most important feature, the addition of other features does measurably improve performance.}
\label{tab:feature_importances}
\end{table*}

\subsection{Alerts, Filtering, Streaming and \skyportal Integration}
After completing processing, each detection is converted to an \emph{Apache} \texttt{Avro} alert, adopting the same data format used by ZTF \citep{ztf_bellm_19}, Gattini \citep{gattini}, and Rubin \citep{Rubin}. The WINTER alerts contain information about both the latest detection and historical detections, as well as the contextual data, the source name, and the latest cutout images. All Avro alerts contain a full schema detailing the data products, and the \winter alerts contain additional information about schema version. These Avro alerts are written to disk. 

Due to the large number of raw WINTER detections (a typical night might produce 20000 detections), substantial additional filtering is required to identify sources of scientific interest. Cuts are applied based on the source metadata, including parameters such as \texttt{rb}/\texttt{xrb} and number of prior detections. Moreover, as the differencing infrastructure is designed to identify transients, additional cuts are applied based on the crossmatching results to remove known stars. Ultimately, approximately $\sim$150 alerts survive these cuts in a given night and are deemed potentially interesting. 

\begin{figure*}
    \centering
    \includegraphics[width=\textwidth]{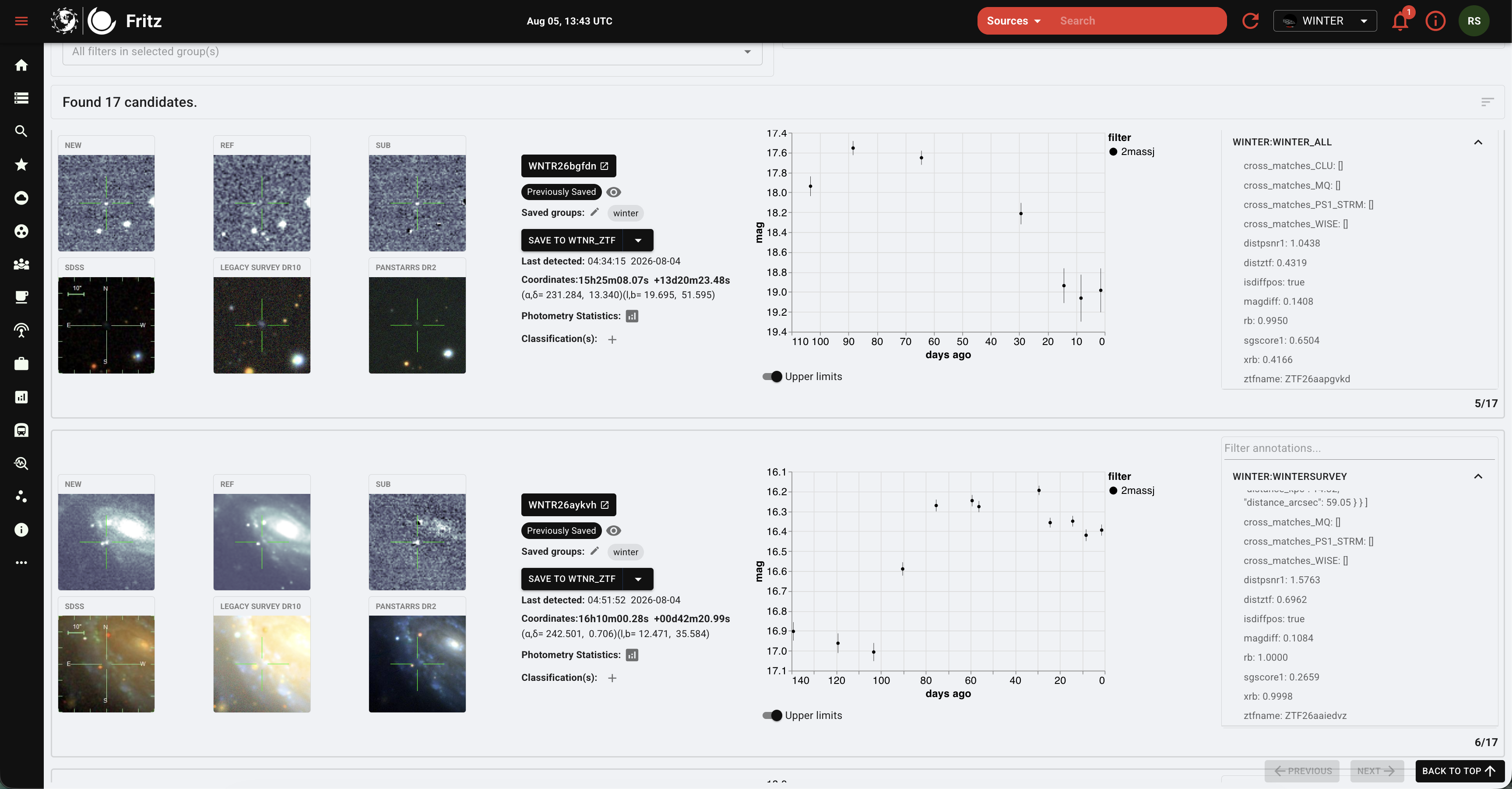}
    \caption{\skyportal scanning page of the WINTER+ZTF filter for the week 2026-07-29 to 2026-08-05, featuring two supernovae (SN\,2026hiu and SN\,2026ejy) that were detected by both \winter and ZTF.}
    \label{fig:skyportal}
\end{figure*}

After applying the cuts, the \avro alerts for the selected sources are transferred to IPAC, which
serves as a broker and distributes them more widely as a \kafka `alert stream' \footnote{\url{https://kafka.apache.org/}}, following the same
model used for ZTF \citep{ztf_bellm_19}. The stream is consumed by \texttt{BOOM} \citep{boom_2026}, a \texttt{Rust}-based real-time brokering system operated at Caltech. \texttt{BOOM} was built to process the ZTF and LSST alert streams jointly at Rubin-era rates, and now also ingests and archives every \winter alert. User-defined filters are executed by \texttt{BOOM} against the ingested alerts as database aggregation queries, and only those alerts which pass a given filter are posted as candidates to the \url{fritz.science} instance of
\skyportal \citep{skyportal_19,skyportal_23} used by the ZTF and \winter collaborations. These
filters can overlap, but each is focused on a particular science case such as `transients in nearby galaxies', `transients with a ZTF crossmatch', or `extragalactic transients'. Human scanners vet the candidates which pass these filters, of which there are typically 120 per night, and bona fide sources are saved for further follow-up.
\skyportal then serves as the interface for coordinating that follow-up, including the submission of new \winter ToO requests via the API described in Section\,\ref{sec:winterapi}. An example of the \winter scanning page is shown in Figure \ref{fig:skyportal}.


\subsection{WINTER Database}
\label{sec:database}
A single \texttt{PostgreSQL} database is used to store metadata about WINTER data, with a table and field structure that is automatically generated by \drp through the \texttt{SQLAlchemy} \python package \citep{sqlalchemy}. The database tables are indexed by spatial position using \texttt{Q3C} \citep{q3c}, increasing the speed of spatial queries and crossmatching.  The database includes separate tables corresponding to images at five stages of processing:

\begin{itemize}
    \item \textbf{Exposure} - The original WINTER camera images, with 6 multi-extension fits frames corresponding to the six individual camera sensors.
    \item \textbf{Raw} - The individual WINTER camera images, after the multi-extension file has been split into six separate images and pixel masking has been applied.
    \item \textbf{Science} - Science-ready WINTER camera images, after detrending and astrometric/photometric calibration has been applied. 
    \item \textbf{Stack} - The stacks of all individual WINTER science images for a given detector, including each image in the dither set.
    \item \textbf{Diff} - The difference images generated for WINTER stack images, after image subtraction against a reference image.
\end{itemize}

Further tables include ones for filters, fields and active survey programs. There are two additional tables relating to transients:

\begin{itemize}
    \item \textbf{Sources} - All objects detected at least once by \winter, including a unique name for each.
    \item \textbf{Candidates} - Every individual detection by \winter. A single source must have one or more detections associated with it.
\end{itemize}

These database tables are used by both the Grafana dashboard (Section \ref{sec:grafana}) and the API (Section \ref{sec:winterapi}). Most \drp interactions with the database come via \processors which simply export metadata to an individual database table. However, \drp also queries the source table to determine whether a named object already exists at the position of each detected transient, and the candidate table to retrieve the complete detection history of sources which have been detected before.  

\subsection{\textit{Grafana} Dashboard and Alerts}
\label{sec:grafana}
\begin{figure*}
    \centering
    \includegraphics[width=\textwidth]{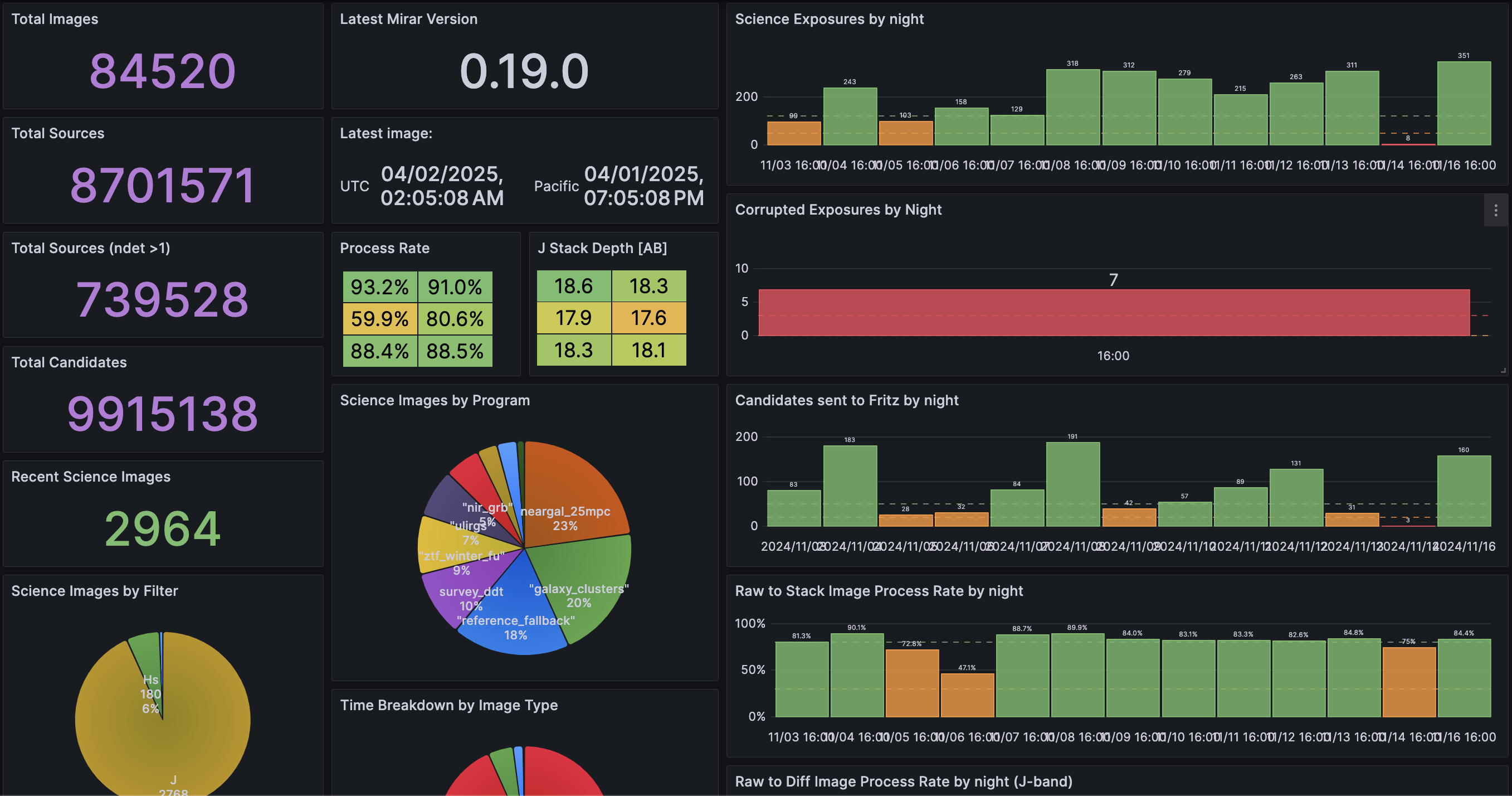}
    \caption{A panel from the WINTER Grafana dashboard showing a summary of recent data and survey performance.}
    \label{fig:grafana}
\end{figure*}

\wdrp performance is monitored through a dedicated \textit{Grafana}\footnote{\url{https://grafana.com}} dashboard, which provides live visualisation of the database. An example of a dashboard page is shown in Figure \ref{fig:grafana}. The dashboard provides processing efficiency, an aggregate log of all recent ToO requests, as well as photometric/astrometric performance, and overall statistics about the time usage of each survey program. The \textit{Grafana} dashboard has been configured to provide automated email alerts for a variety of data problems including a lack of new images for a given night, an abnormally low processing rate, or images arriving with corrupted or missing header information. These notifications help to promptly identify data-taking issues that may arise, so they can be resolved promptly.  

\subsection{Image access and target submission API}
\label{sec:winterapi}
WINTER observation requests and data access are handled by a dedicated API server, hosted on a dedicated winter server at Caltech. The API is built using the \texttt{fastapi} \python framework \citep{Ramirez_FastAPI}, and includes automated generation of OpenAPI-compliant metadata. These metadata are used to create an interactive documentation of the API using the \texttt{Swagger UI}. The documentation is hosted on a web page for WINTER users\footnote{\url{http://winter.caltech.edu:82/docs}}. The API ensures that minimal human intervention is required to support user interaction, and enables Target-of-Opportunity (ToO) and other user-generated scheduling requests to be handled with very low latency. All requests for observations with WINTER require a validity window and priority, and are passed on with this information as `ToO requests' to the WINTER scheduler.

A dedicated python package, \wintertoo\footnote{\url{https://github.com/winter-telescope/wintertoo}}, provides a single shared framework for these ToO requests. \wintertoo is used by the WINTER API server, as well as the WINTER camera software and \drp. \wintertoo contains a standardized format for ToO requests as well as a set of functions to ensure that the ToO requests are valid. These include checking that the target is observable within the requested window above the requested airmass limit, that the program has sufficient unexpired time, as well as more low-level checks such as requiring the requested filters to actually be one of those mounted on the camera. Once validated by the WINTER API server, ToO requests are copied to the WINTER computer at Palomar. After any current observation has been executed, these new ToO requests are read by the WINTER Supervisor Program (WSP)\footnote{\url{https://github.com/winter-telescope/observatory}}, which manages the observatory operations \citep{wsp_24}. Additional validation checks are repeated at this stage, and valid requests are added to the queue. The API server monitors the status of these requests, which can be either `QUEUED' or `ATTEMPTED'. To prevent the scheduler getting stuck attempting a problematic target, each individual ToO observation is attempted no more than three times on a given night. Unsuccessful ToO requests, for example due to hardware failures that prevent data acquisition, are not charged to respective programs. A WINTER log provides a complete list of all data that was successfully taken and these observations are charged to the respective programs. 

Once the data has been processed and updated in the \wdrp database, it becomes available for users to access. This process is also handled by the winter API server. Users can query the API to find available images for a given program at the five stages of data processing described in Section \ref{sec:database}. Users can also query for the complete set of Avro alerts generated for a given WINTER difference image. These queries can either provide all images associated to a program, or users can select subsets based on target name and/or observation date. Users can then perform a separate API request to download a list of these images as a ZIP file. For security, each user requires unique user credentials to interact with the API, and each WINTER program has an associated set of unique program credentials. Both sets of credentials are required to request observations for a program or access data from a program. 

Users primarily interact with the WINTER API server through a separate python package, \winterapi\footnote{\url{https://github.com/winter-telescope/winterapi}}. This package has simple wrapper functions for interacting with the WINTER API server, as well as example Jupyter notebooks illustrating the various features. Beyond allowing users to manually interact with the API server in a simple way, \winterapi is designed to be easily imported as a dependency into user-built \python scripts, simplifying automated interactions with the API. \winterapi is an essential component of the software infrastructure supporting automated GRB follow-up with WINTER, as well as MMA follow-up of gravitational wave and neutrino events. Alternative API access is available through \skyportal \citep{skyportal_19,skyportal_23}, the {Swagger UI} documentation page or directly via the server. 

\section{On-sky performance}
We now quantify the on-sky performance of the WINTER detectors and the pipeline. We primarily focus on metrics from J-band observations, which comprise majority of \winter's operations. 
\label{sec:performance}
\subsection{Processing Rates and Latency}
The average successful processing rate from raws to stacks is $\approx85$\% across all detectors during periods when no detectors are affected by hardware failures. Processing failures primarily occur in images affected by clouds or bad weather, or images taken during twilight with high background counts. The average stack-to-difference image rate from routine operations is $\approx70$\%. Failures include fields that were not covered by the UKIRT surveys (usually $\delta \geq 60$\,deg, but also some southern fields with $\delta<-25$\,deg), and stacked images affected by weather with shallow depths that cannot be matched to the deeper UKIRT reference images. We note that during periods of good weather with observations limited to regions with reference coverage (e.g. GW followup of S250206dm, see Sec. \ref{sec:early_results}) stacks-to-difference rates are much higher ($\geq90$\%).

As \winter requires dark images to be taken at midnight, the processing of the data begins at 1 am Pacific time every night, and the full night's data processing typically finishes before 9 am Pacific. A daily email notifies the WINTER users of the completion of the pipeline. 

\subsection{Sensitivity}
\begin{figure}
    \centering
    \includegraphics[width=\linewidth]{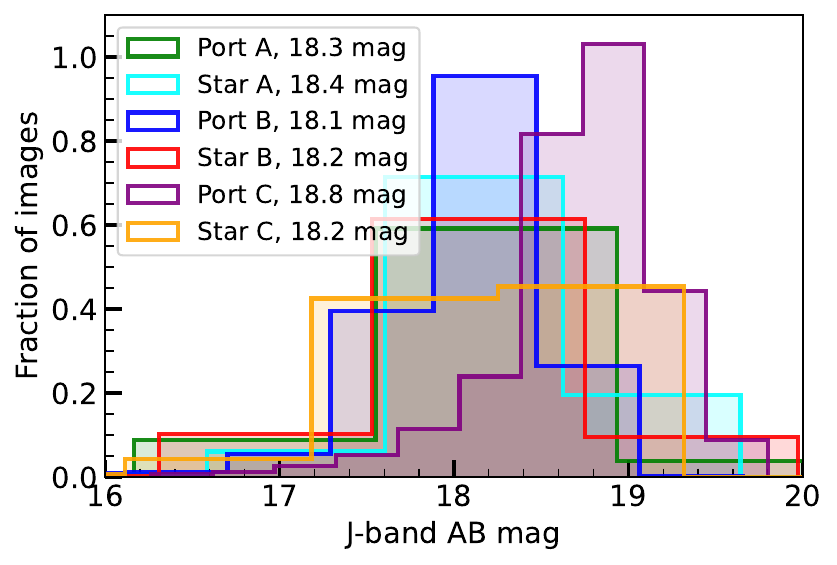}
    \caption{Distribution of the 5-$\sigma$ limiting magnitudes for each of the six WINTER sensors, for data taken during the calendar year 2024 with the default survey total integration time of 960 seconds. The median values of the limited magnitudes range from 18.1 to 18.8\,mag (AB), and are listed in the legend.}
    \label{fig:maglims_j}
\end{figure}

Figure \ref{fig:maglims_j} shows the distribution of 5$\sigma$ \emph{J-}band limiting magnitudes for the default 960 second exposures (8 dithers, 120 seconds each) of the routine WINTER survey taken from all images taken during the calendar year 2024. The median limiting magnitudes range from 18.8\,mag (AB) on the most sensitive detector Port C to $\sim$18.1\,mag (AB) on the least sensitive detectors. For similar total integration times of 960 seconds, \winter reaches a 5-$\sigma$ limiting magnitude of $17.6-18.4$\,mag (AB) in the y-band and $16.5-17.5$\,mag (AB) in the Hs-band, depending on the detector.

\subsection{Improvements from non-linearity corrections}
\begin{figure*}[!hbt]
    \centering
    \includegraphics[width=0.5\textwidth]{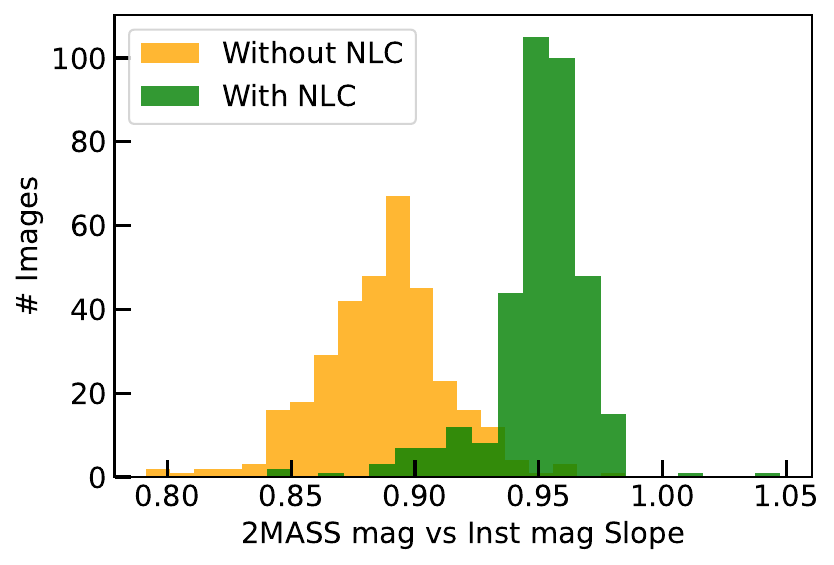}\includegraphics[width=0.5\textwidth]{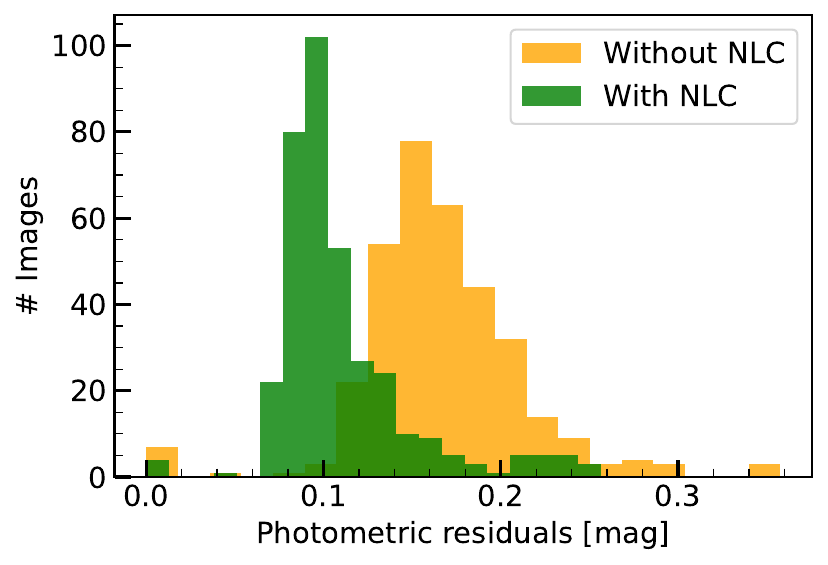}
    \caption{Improvements in performance after applying the current version of non-linearity corrections (NLC) on \winter's deepest channel (Port C). \emph{Left:} Distribution of slopes obtained by fitting a linear function to the instrumental magnitudes as a function of reference magnitudes. It is clear that the slopes after applying the corrections are closer to perfect linearity (1) than the slopes without NLC. \emph{Right:} Photometric residuals after NLC are lower than those without NLC. Efforts are underway to further characterize the NLC of the WINTER sensors.}
    \label{fig:nlc_effects}
\end{figure*}

We now demonstrate the improved photometric accuracy due to NLCs as described in Section\,\ref{sec:nlc}. To assess \winter's photometric accuracy, we use \emph{J-}band data taken during a ten-day period of stable operations from UT 2025-02-15 to 2025-02-25. We quantify the extent of detector non-linearity using the slope between instrumental magnitudes and 2MASS reference magnitudes of the stacked images, measured by performing a linear fit between these quantities. A value of the slope of unity implies perfect linearity, while deviations indicate the level of non-linearity. Figure \ref{fig:nlc_effects} (left panel) shows distributions of these slope values measured for all images in the ten-day period from detector Port C, with and without applying non-linearity corrections. We find that using non-linearity corrections yields slopes closer to unity than the stacks without corrections (median value 0.96 vs 0.88). Figure\,\ref{fig:nlc_effects} (right panel) shows the distribution of the photometric residual, calculated for each stacked image as the median RMS offset between the instrumental magnitudes and 2MASS magnitudes for bright stars ($12<m_{J}<15$), with and without NLC. We find significantly lower photometric residuals with a median residual value $\approx0.10$\,mag in stacks with NLC compared to 0.16\,mag for stacks without. We also find substantially fewer bad pixels in the dark-calibrated images than before. These results demonstrate the efficacy of the NLC in improving the detector response. 

\subsection{Photometric precision}
\label{sec:performance_phot}
\begin{figure*}
    \centering
    \includegraphics[width=0.5\textwidth]{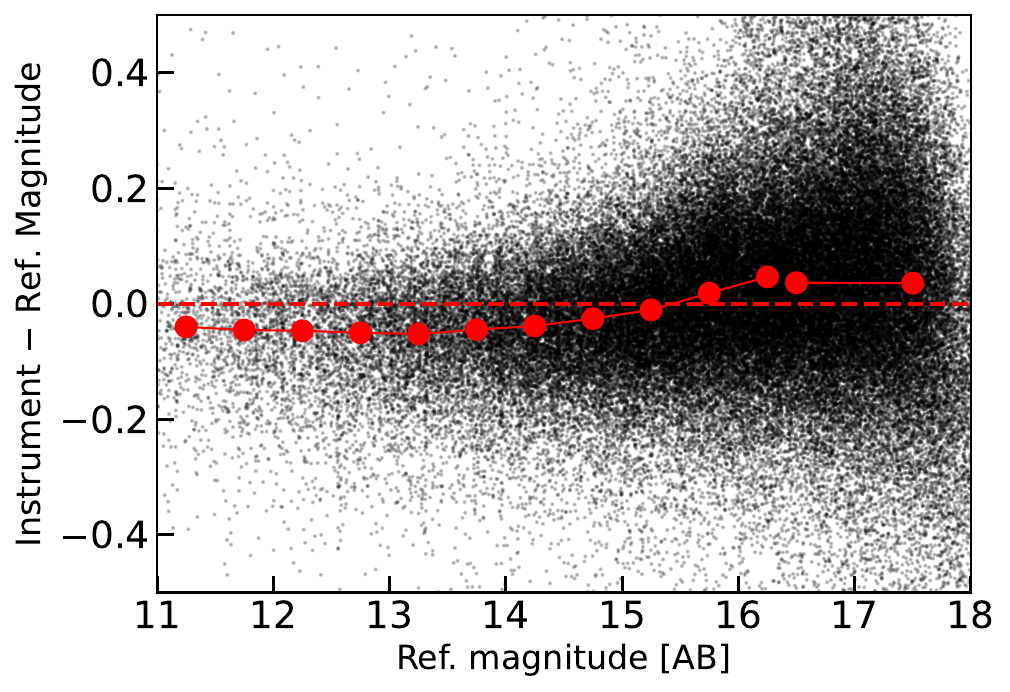}\includegraphics[width=0.5\textwidth]{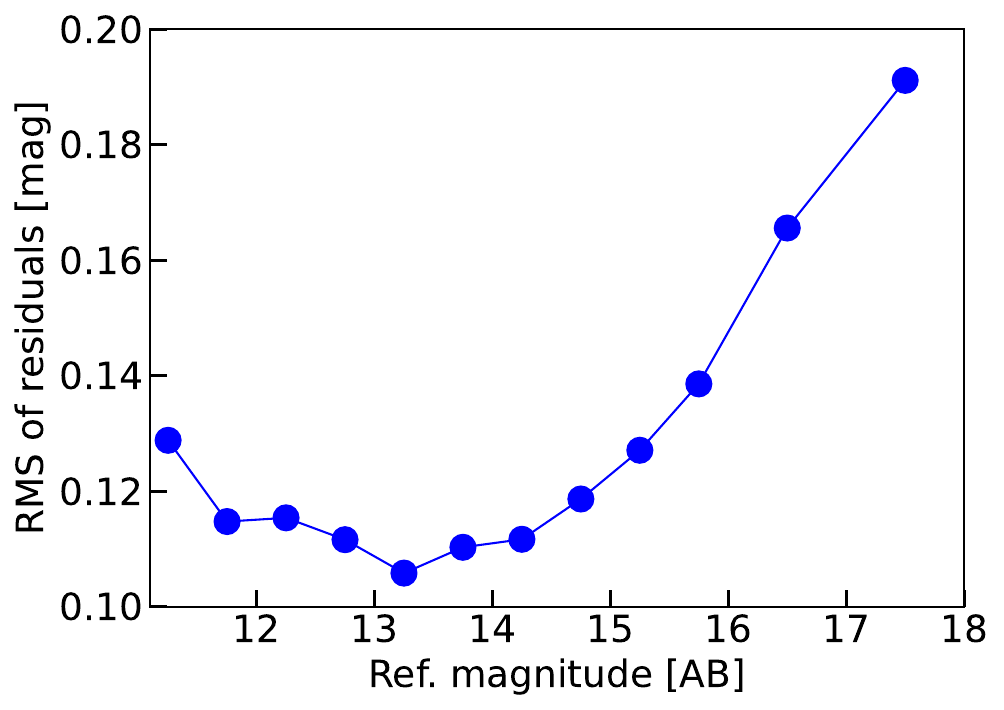}
    \caption{\emph{Left}: Photometric scatter, computed as the difference between measured and reference magnitudes of sources detected in Port C \winter coadded stacks (see text for details). The red dashed line marks the line of zero scatter. The red dots mark median residual values in 0.5-magnitude bins, indicating a residual non-linearity in the response at the $\approx0.04$\,mag level. \emph{Right:} RMS of the photometric scatter, measured in 0.5-magnitude bins as a function of source magnitude for Port C \winter stacks. This reveals a systematic uncertainty floor of $\approx0.11$\,mag even for bright sources ($12<m_J<15$).}
    \label{fig:phot_residual_mean_rms}
\end{figure*}

For all \emph{J-}band data taken during the ten-day period described above, we calculate the offsets between the measured instrumental magnitude and reference magnitudes. We use the 2MASS J-band magnitudes as reference for sources with $12 < m_J < 16$. For fainter sources, we use \texttt{SExtractor} to perform point-spread-function photometry on the UKIRT reference image, and use the source catalogs as reference catalogs. Figure\,\ref{fig:phot_residual_mean_rms} (left panel) shows the difference between instrumental and reference magnitudes, as a function of reference magnitude for all sources detected on Port C during this period. Red dots indicate the median offsets in 0.5-magnitude bins. We find that the median differences are offset from zero by $\approx-0.04$ mag at the bright end ($m_J=12$) and by $\approx+0.04$\,mag at the faint end ($m_J=18$) - indicating a residual non-linearity in the system at the $\approx5\%$ level. Similar trends are observed for the other five detectors.

\begin{figure}
    \centering
    \includegraphics[width=0.5\textwidth]{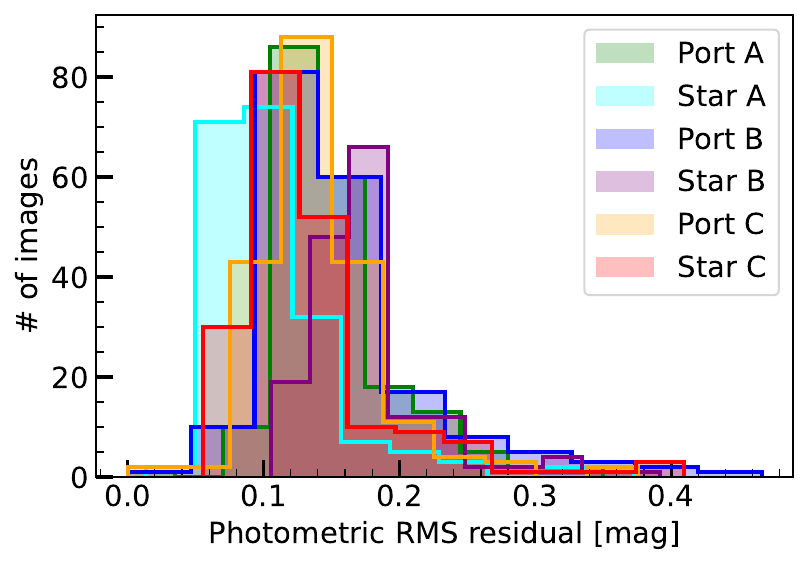}
    \caption{Distribution of residuals computed by measuring the RMS variations of the difference between instrumental magnitude and reference magnitudes of bright stars in photometrically calibrated \winter stacks. Different colors show the six different \winter detectors. The median of the residuals range from 0.09-0.18\,mag, which set the systematic uncertainty floor for WINTER photometric measurements.}
    \label{fig:photcal_residuals}
\end{figure}

Figure\,\ref{fig:phot_residual_mean_rms} (right panel) shows the root-mean-square (RMS) of the instrumental and reference magnitude differences calculated in 0.5\,mag bins, as a function of reference magnitude. We find RMS values of $\approx0.2$\,mag at the faint end ($m_J\approx18$). However, we find relatively high RMS values of $\approx0.11$\,mag even for brighter sources ($12<m_J<16$), which is largely independent of source magnitude -- indicating that this is an uncertainty floor for the photometric performance of detector Port C. We quantify this uncertainty floor for every image by computing the RMS of the instrument and reference magnitude difference for bright sources with $12<m_J<16$. Figure \ref{fig:photcal_residuals} shows the distribution of these values for every \winter detector. The median photometric residuals range from the lowest value of 0.09\,mag on Star\,A to the highest value of 0.18\,mag on Star\,B. This uncertainty floor is likely due to dark current fluctuations ($\sim$5\% relative to sky counts as described in Section\,\ref{sec:dark_calibration}), as well as sub-optimal flat-fielding strategies, non-linearity effects, pick-up noise from electronic components -- each of which contribute a few percent to the calibration errors. To account for this systematic uncertainty in the reported photometry, we compute, for each source, the RMS offset between the detected and 2MASS magnitudes of all sources within a 0.5 mag bin centered on the source's magnitude, and add it in quadrature to the statistical uncertainty.


\subsection{Image subtraction and transient recovery}
\begin{figure*}
    \centering
    \includegraphics[width=0.5\textwidth]{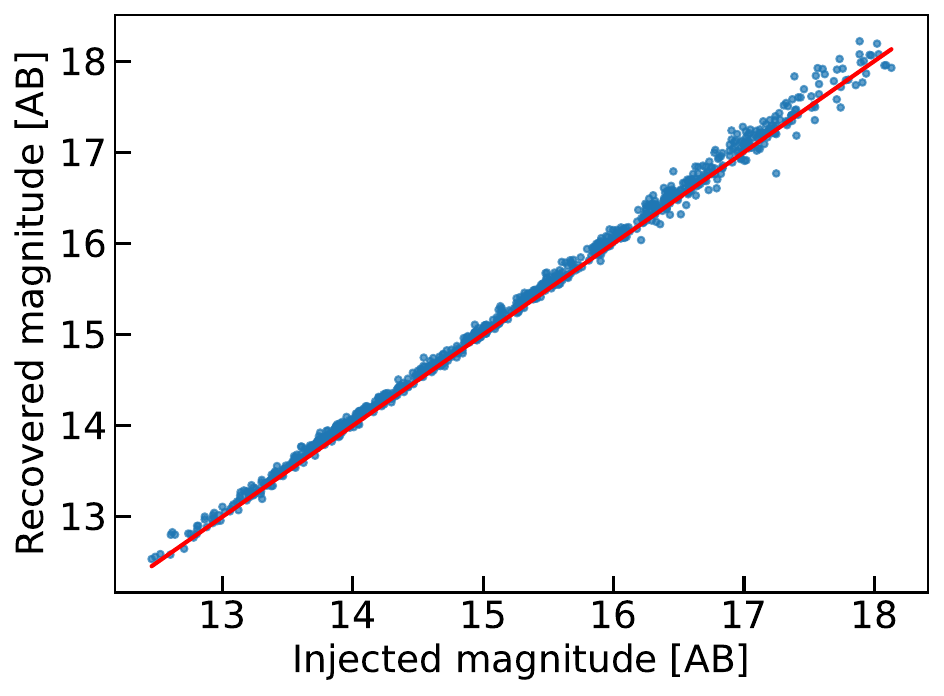}\includegraphics[width=0.5\textwidth]{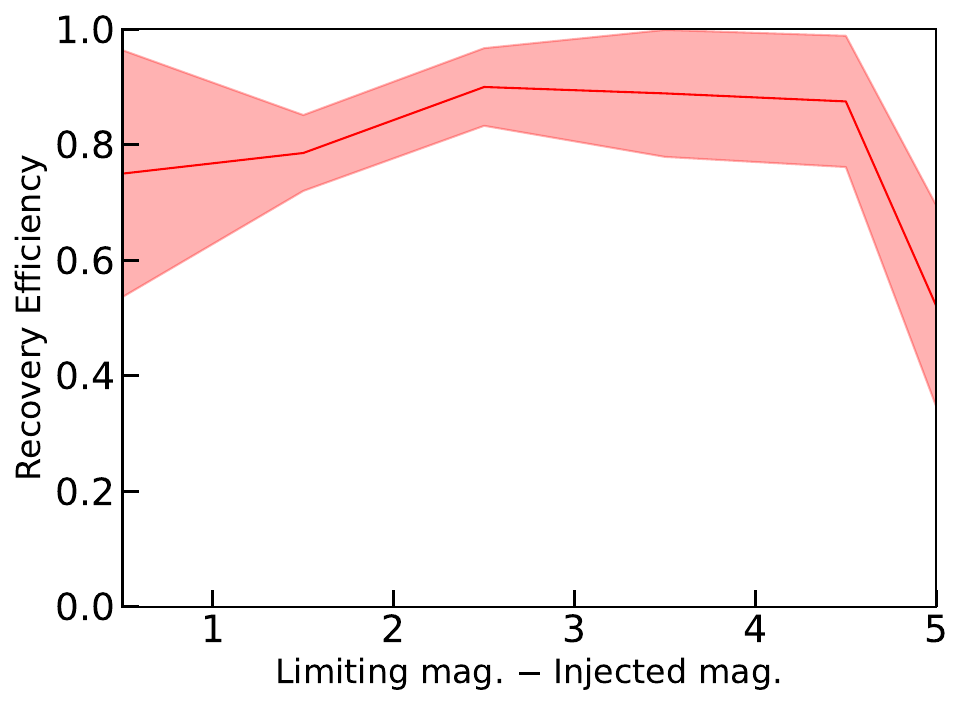}
    \caption{\emph{Left:} Measured PSF magnitude from the difference image as a function of injected magnitude for all fake sources recovered by the difference imaging pipeline. \emph{Right:} Recovery efficiency, computed as the ratio of the number of recovered candidates to injected candidates as a function of source brightness relative the limiting magnitude of the image. The red line shows the median recovery efficiency over eleven \winter stacks from Port C, while the shaded region denotes 1-$\sigma$ uncertainties.}
    \label{fig:recovery_efficiency}
\end{figure*}

We quantify the performance of the image subtraction pipeline by performing fake-source injection and recovery tests. We injected a total of 1100 sources in eleven stacked images from the Port C detector (100 sources per image) from the test period using their measured PSFs. The brightness of the sources is uniformly distributed between 0 and 5 magnitudes above the limiting magnitude of the image. The spatial locations of the injected sources were chosen randomly over the entire image. We ran the default image subtraction pipeline end-to-end on these sources and identified those that were recovered as candidate transients. We find a total of 894 recovered transients, yielding an overall recovery rate of $\approx82\%$. Figure\,\ref{fig:recovery_efficiency} (right panel) shows the recovery efficiency as a function of the difference between source magnitude and the limiting magnitude of the image. The recovery efficiency is generally high ($\gtrsim90\%$) for sources 2--4 magnitudes brighter than the limiting magnitude, drops to $\approx70\%$ for sources within 1 magnitude of the limiting magnitude, and $\approx$50\% for sources 5 magnitudes brighter than the limiting magnitude. The faint end drop-off is due to increased noise near the detection threshold of the image, while the brighter end drop-off is due to masking of bright and saturated sources in the pipeline. Most of the sources that were not recovered belonged primarily to these two categories, while the remaining sources were not recovered because they were close to masked regions on the image or in close proximity of bright stars. The measured PSF magnitudes of the recovered sources from the difference images are compared with their injected magnitudes in Figure\,\ref{fig:recovery_efficiency} (left panel), which shows no systematic offsets between the recovered and injected magnitudes across the range probed by these images. 

\section{Science results from \winter} 
\label{sec:early_results}
Since first light in June 2023, \winter has been operating robotically, executing both long-term surveys and target-of-opportunity (ToO) observations. Long-term surveys include surveys of nearby galaxies, the Galactic plane, and the WINTER reference building survey. ToO observations include a dedicated program for electromagnetic followup of gravitational waves (EMGW), followup of gamma-ray bursts (GRBs) and fast X-ray transients (FXTs), followup of high-energy neutrino alerts and followup of the reddest ZTF transients. We list examples from these observing programs below. 
\subsection{Survey science}
With \winter, we are conducting a J-band survey comprising five pointings covering the M31 group observed at approximately weekly cadence. In 2024, we identified the source WNTR\,23bzdiq (Figure\,\ref{fig:science_demos}) in this survey: a slowly evolving transient in M31 that resembled expectations for the onset of common-envelope evolution in a binary system with an asymptotic giant branch primary star \citep{Karambelkar2025}. In November 2025, the slow variations transitioned to a rapid $\approx10\times$ brightening \citep{Taguchi2025ATel}, with properties resembling Luminous Red Novae (LRNe) -- a class of eruptions originating from stellar mergers \citep{Kaminski2026}, making WNTR\,23bzdiq the first LRN discovered during its pre-eruption phase \citep{karambelkar_cee}.

With \winter, we are also conducting a survey of regions in the plane of the Milky Way with galactic latitude $|b|<2 \mathrm{^{o}}$ and galactic longitude $15<l<50$. This region was chosen to overlap with the footprint of the planned Roman Galactic Plane Survey (GPS) that is visible from the northern hemisphere. In the 2025 Galactic observing season, we identified WNTR\,25fhdoi: a slow nova in this survey (\citealt{Karambelkar2025_atel}, Figure\,\ref{fig:science_demos}). This eruption is still ongoing in 2026, lasting for over three hundred days, signaling an exceptionally long duration for this unusual nova. In the 2026 Galactic season, we independently identified the reddened classical nova AT\,2026rdg in this survey. This IR-bright nova was saturated in the \winter images, and was recovered retroactively. 

\subsection{ToO science}
\subsubsection{NIR follow-up of known transients}
\begin{figure*}
    \centering
    \includegraphics[width=0.49\textwidth]{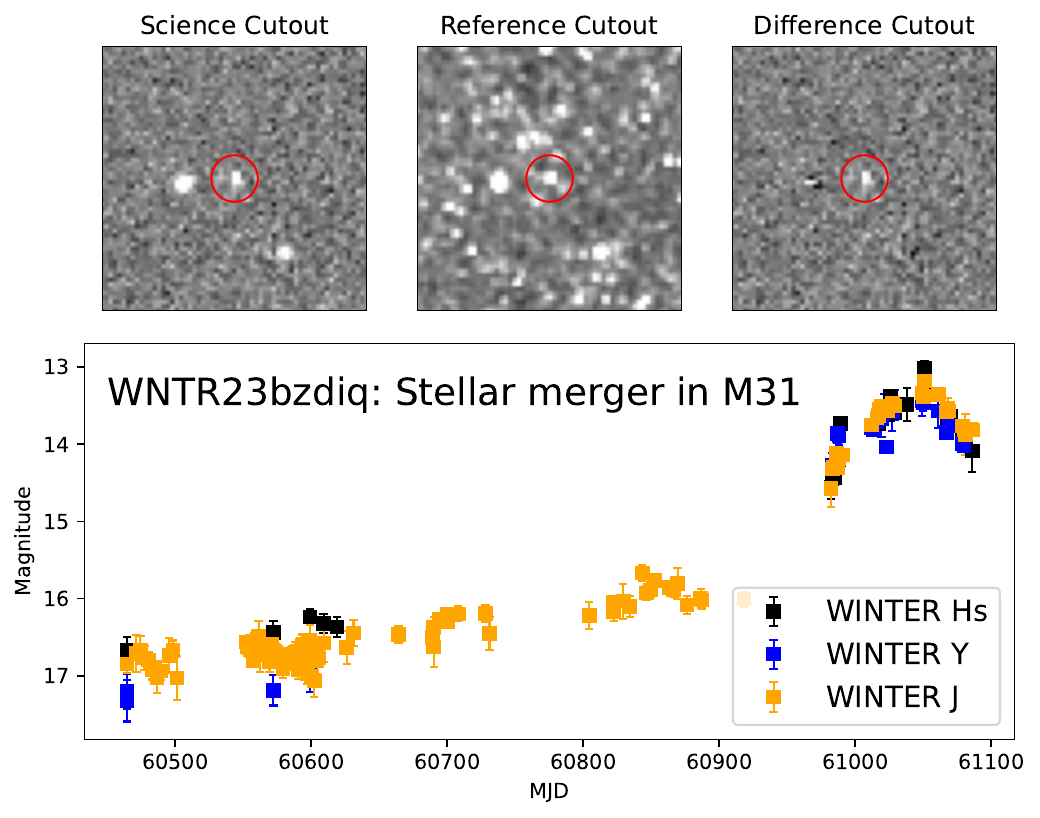}
    \includegraphics[width=0.49\textwidth]{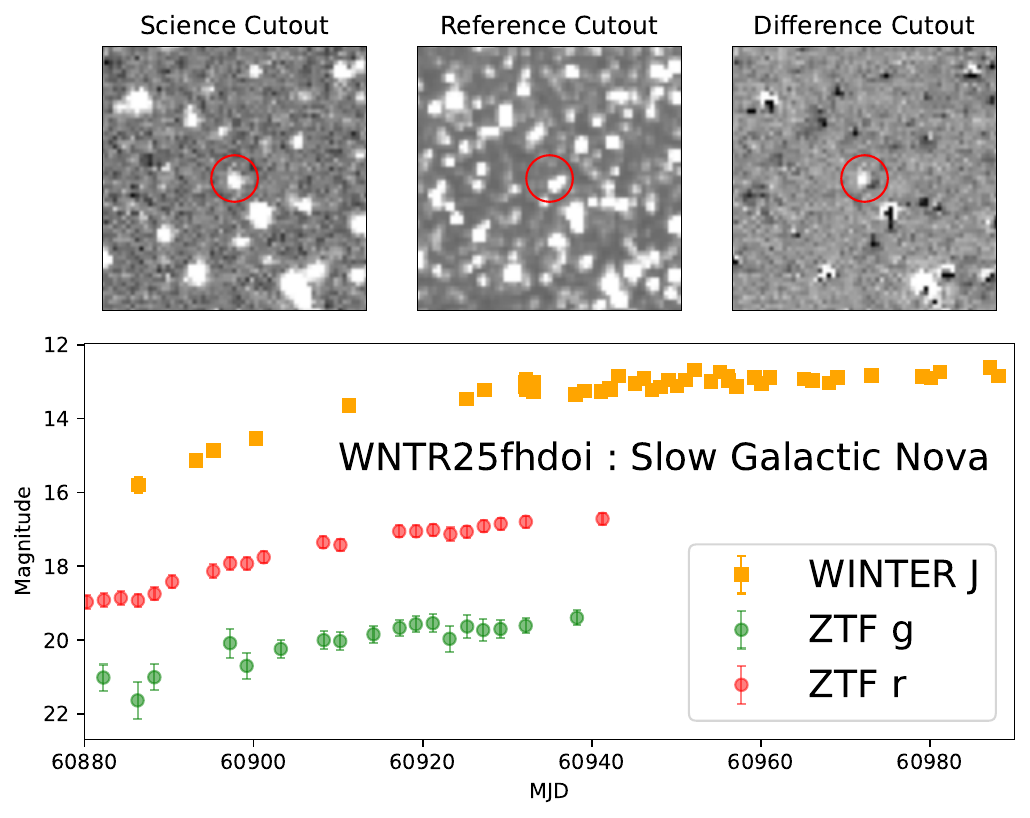}\\
    \includegraphics[width=0.49\textwidth]{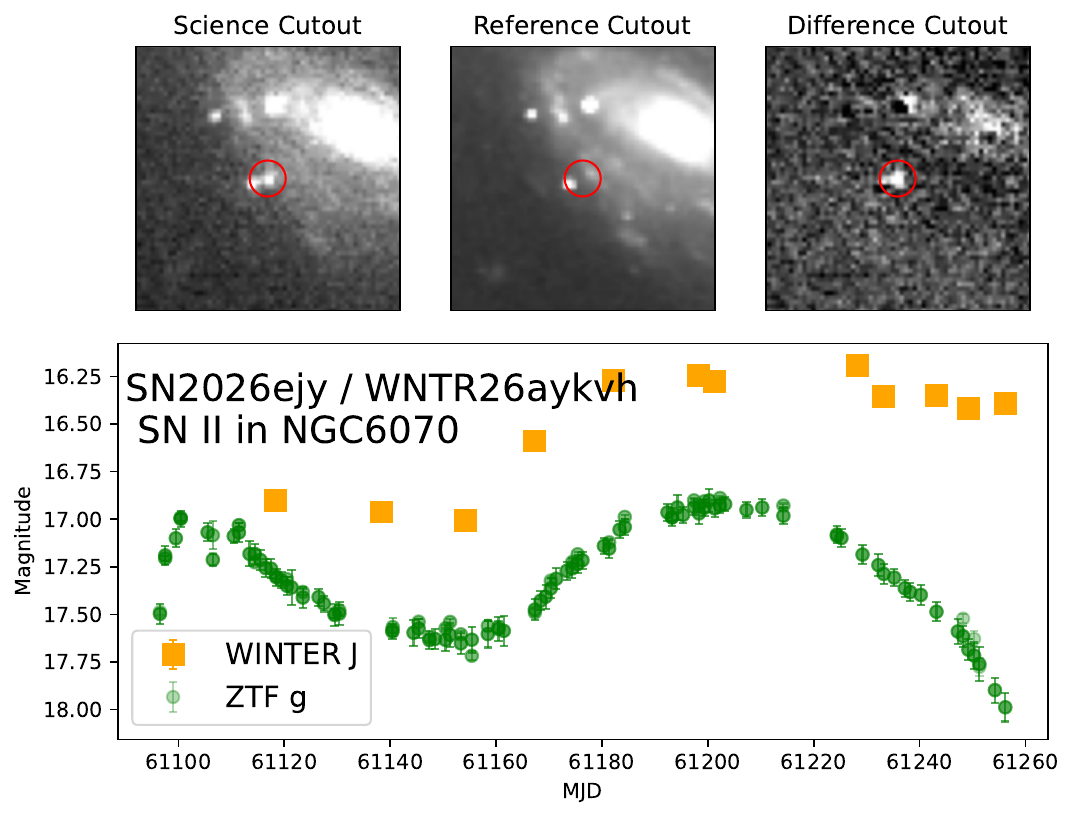}\includegraphics[width=0.49\textwidth]{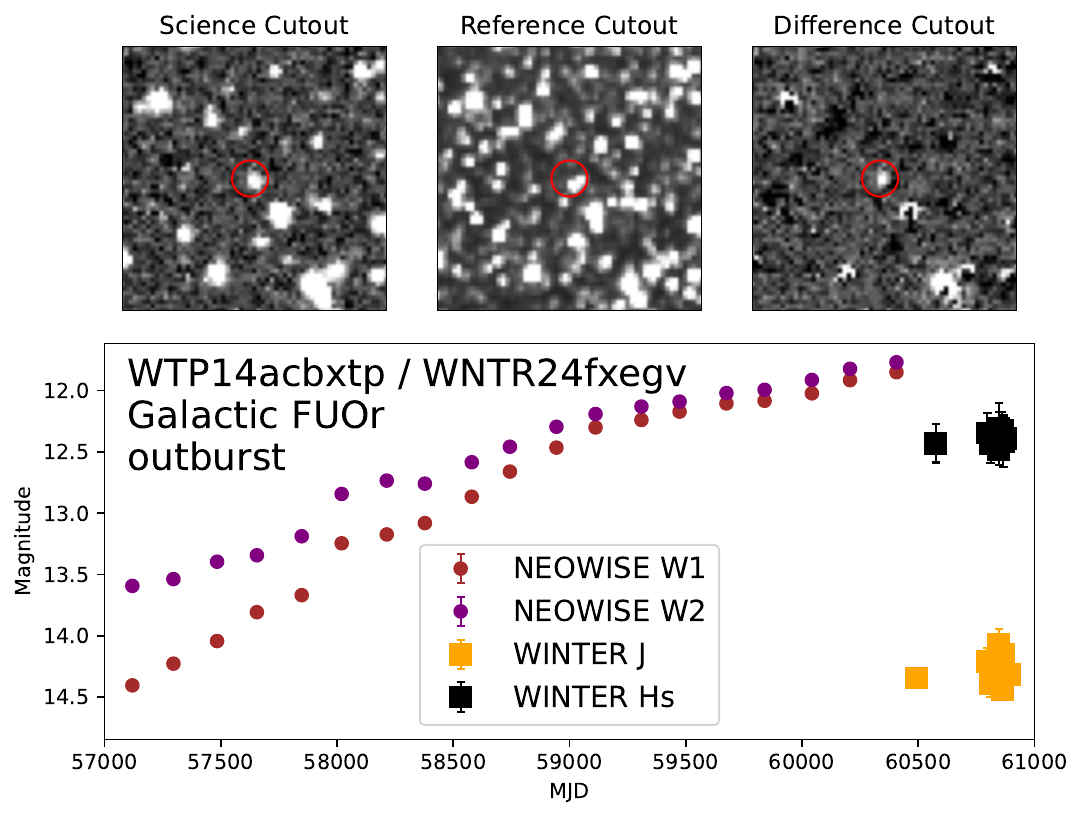}
    \caption{Examples of Galactic and extragalactic transients identified (top) or followed-up (bottom) with \winter (see Section\,\ref{sec:early_results} for details).}
    \label{fig:science_demos}
\end{figure*}

A dedicated \winter program systematically follows up red extragalactic transients identified by ZTF in the \emph{J-}band, to identify signs of dust formation in them. Figure \ref{fig:science_demos} shows an example of SN\,2026ejy followed up with \winter. So far, \winter has obtained \emph{J}-band observations for 482 ZTF transients \citep{rats_paper}. Of those 482 objects, 317 had reference image coverage and produced a difference image. Of the 317 objects with difference images, 220 were detected, and an additional 72 objects (without reference images) were detected in an unsubtracted image stack. In addition to extragalactic transients, WINTER ToO programs also observe Galactic transients identified by surveys such as ZTF and NEOWISE \citep{Mainzer2014, De2023Nature}, and has, e.g., contributed to the IR characterization of highly extinguished FU-Ori outbursts \citep[][see also Figure\,\ref{fig:science_demos}]{frostig_26} and a newly discovered recurrent nova in the Milky Way \citep{frostig_nova}. 

\subsubsection{Multi-messenger followup}
One of \winter's main science goals is IR followup of GW alerts to search for associated kilonovae \citep{Frostig:2022}. During LVK's O4 observing run, \winter followed up the NSBH GW alert S250206dm, which had a median distance of $\approx$373\,Mpc and a 50\% sky-localization area of 38\,sq.\,deg. \citealt{s250206dm} presents the results of WINTER follow-up for this trigger, which covered a total of $\approx$43\% of the integrated probability of the sky localization and did not find any compelling counterpart candidate. While the \winter upper limits are not sensitive enough to constrain kilonova models at the distance of this event, our search demonstrates \winter's capabilities to tile GW localizations, and highlights the promise of NIR EMGW followup. 

In addition to EMGW followup, \winter has a dedicated program for automated observations of GRBs detected by the \textit{Neil Gehrels Swift Observatory} \citep{gehrels_04}, the Space-based multi-band astronomical Variable Objects Monitor (SVOM) mission \citep{svom}, and fast X-ray transients (FXTs) detected by Einstein Probe \citep{einstein_probe}. The localization regions (typically 1-3\,arcmin) of these sources are small enough to be covered by the most sensitive \winter detector (Port C). So far, we have reported \emph{J-}band observations for 61 GRBs and EP sources (e.g. \citealt{Mo2024_GCN, Ahumada2024_winter_ep, Karambelkar2024_GCN}), including 12 detections. The automated triggering ensures \winter can begin observations within a few minutes of the event being observable from Palomar. We continue to automatically trigger these alerts, making use of the WINTER API infrastructure introduced in Section \ref{sec:winterapi}.

\winter also has a dedicated program for automated triggering of high-energy neutrino alerts from IceCube \citep{icecube_alerts,icecube_alerts_v2}. As an example, during the follow-up campaign of the neutrino alert IC\,240105A, \winter covered the blazar PKS\,0446+11 that lies within the localization region and was reported to undergo a flare possibly coincident with the neutrino alert. We detect the flaring of this blazar in our \winter observations, with the \emph{J}-band magnitude being 2\,mag brighter than the optical \emph{r-}band \citep{Stein2024_GCN_winter_neutrino}. We have since automated the triggering procedure, ensuring low-latency observations can be obtained for all well-localised neutrino events. 


\section{Summary and way forward} \label{sec:summary}
In this paper, we present the data reduction and transient detection pipeline for the \winter near-infrared time-domain surveyor. The \wdrp pipeline was implemented using \texttt{mirar}: a modular, open-source \python-based framework designed to process images from any telescope and find transients in them. \texttt{mirar}'s modular approach to the task enables users to plug in any telescope and go from raw frames to \avro alerts, by reusing \processors for performing tasks such as calibrations, resampling, stacking, image subtraction, and source detection. \texttt{mirar} has built-in \processors for widely-used astronomy software such as \texttt{astrometry.net}, \texttt{Sextractor}, \texttt{Scamp}, \texttt{SWarp}, and also supports interactions with the data visualization interface \skyportal. \texttt{mirar} is open-source and heavily tested, and its modular design means that additional capabilities can be easily added as needed.

The \wdrp implemented in \texttt{mirar} processes images from the WINTER surveyor at Palomar Observatory to identify transients in them using template images from the UKIRT public surveys. We quantify the photometric performance of \winter using year-long observations from 2024, and find 5$\sigma$ limiting magnitudes ranging from $\approx$18.1--18.8\,mag (AB) on the six detectors for standard total integration time of 960 seconds. The limiting magnitudes range from $\approx$17.6--18.4\,mag (AB) in the y-band, and 16.5--17.5\,mag (AB) in the Hs-band. We find a systematic uncertainty floor ranging from 0.09-0.18\,mag (depending on the detector) on WINTER photometry, likely due to residual non-linearity effects and dark current variations within the detectors. We conduct fake-source injection tests for the image differencing pipeline and find a candidate recovery rate $\approx90\%$. We present initial science results from \winter operations, including the early identification of a stellar merger in M31, identification of classical novae and dust-enshrouded young-star outbursts in the Milky Way, NIR followup of an NSBH gravitational wave candidate, automated followup of several GRBs, FXTs, and neutrinos, and NIR followup of several known transients. \winter continues to survey the Milky Way and nearby galaxies to search for dusty transients, and conduct automated ToO followup of a wide variety of transients and multi-messenger alerts. These results from \winter demonstrate the efficacy of low-cost InGaAs detectors for time-domain explorations of the dynamic infrared sky.

Looking ahead, \winter is part of the emerging landscape of new and upcoming ground-based NIR time-domain surveys such as DREAMS \citep{Soon2022_dreams}, and Cryoscope \citep{Kasliwal2025_cryoscope}. Together, these surveys will systematically explore the dynamic infrared sky and set the stage for more sensitive searches with the imminent \emph{Nancy Grace Roman Space Telescope}.  

\section{Acknowledgments}
WINTER’s construction and early operations are made possible by the National Science Foundation under MRI grant number AST-1828470. Significant support for WINTER also comes from the California Institute of Technology, the Caltech Optical Observatories, the Bruno Rossi Fund of the MIT Kavli Institute for Astrophysics and Space Research, the David and Lucile Packard Foundation, and the MIT Department of Physics and School of Science. 
The Gordon and Betty Moore Foundation, through both the Data-Driven Investigator Program and a dedicated grant, provided critical funding for SkyPortal. 
We acknowledge the contribution of high-school students A. Drake, N. Lam and S. Sutanto to this work through Caltech's Summer Research Connection (SRC) program.
VK was supported by NASA through the NASA Hubble Fellowship grant \#HST-HF2-51578.001-A awarded by the Space Telescope Science Institute, which is operated by the Association of Universities for Research in Astronomy, Inc., for NASA, under contract NAS5-26555. VK also acknowledges the hospitality of the CCA-Flatiron Institute. 
D.F.'s contribution to this material is based upon work supported by the National Science Foundation under Award No. AST-2401779. This research award is partially funded by a generous gift of Charles Simonyi to the NSF Division of Astronomical Sciences. The award is made in recognition of significant contributions to Rubin Observatory’s Legacy Survey of Space and Time. 
S.H. thanks the LSST-DA Data Science Fellowship Program, which is funded by LSST-DA, the Brinson Foundation, the WoodNext Foundation, and the Research Corporation for Science Advancement Foundation; her participation in the program has benefited this work. 
M.W.C. acknowledges support from the National Science Foundation with grant numbers PHY-2117997, PHY-2308862 and PHY-2409481.
G.M. is supported by the Brinson Foundation through the Brinson Prize Fellowship Program.
J.S's contribution to this work was supported partially by the Australian Government through the Australian Research Council's Linkage Infrastructure, Equipment and Facilities funding scheme (project LE230100063). 
Analysis of WISE data was supported by the National Aeronautics and Space Administration through ADAP grant number 80NSSC24K0663. 

\bibliography{myreferences}{}
\bibliographystyle{aasjournal}

\end{document}